\documentclass[showpacs, pra,onecolumn,preprintnumbers ,amsmath, amssymb, superscriptaddress, aps]{revtex4-2}
\pdfoutput=1

\usepackage{dsfont}
\usepackage{float}
\usepackage{array}
\usepackage{slashed,bbold}

\usepackage{amsmath,amssymb,bm} 
\usepackage{graphicx}
\graphicspath{{Figures/}}

\usepackage[tight]{subfigure} 

\usepackage{color} 
\usepackage[papersize={8.5in,11in}]{geometry}

\usepackage{color}
\definecolor{darkblue}{rgb}{0.,0.,0.4}
\definecolor{darkred}{rgb}{0.5,0.,0.}
\definecolor{BlueViolet}{RGB}{138,43,226}
\definecolor{SkyBlue}{RGB}{30,144,255}
\definecolor{DarkGreen}{RGB}{0,100,0}
\usepackage[pdftex,colorlinks=true,linkcolor=darkblue,citecolor=blue,urlcolor=darkred]{hyperref}

\def \be{\begin{align}}
\def \ee{\end{align}}
\def \bea{\begin{eqnarray}}
\def \eea{\end{eqnarray}}

\begin{document}

	\title{ Electron scattering of  mass-inverted in graphene quantum dots  }
	\date{\today}
	\author{Fatima Belokda}
	\affiliation{Laboratory of Theoretical Physics, Faculty of Sciences, Choua\"ib Doukkali University, PO Box 20, 24000 El Jadida, Morocco}
		\author{Ahmed Jellal}
	\email{a.jellal@ucd.ac.ma}
	\affiliation{Laboratory of Theoretical Physics, Faculty of Sciences, Choua\"ib Doukkali University, PO Box 20, 24000 El Jadida, Morocco}
	\affiliation{Canadian Quantum Research Center, 204-3002 32 Ave Vernon,  BC V1T 2L7, Canada}
		\author{El Houssine Atmani}
	\affiliation{Laboratory of Nanostructures and Advanced Materials, Mechanics and Thermofluids, FST Mohammedia,
		Hassan II University, Casablanca, Morocco}


	\begin{abstract}	
	We study the scattering of Dirac electrons of
	 circular  graphene quantum dot with mass-inverted subject to an 
	 electrostatic potential. 
	 The obtained solutions of energy spectrum 
	 are used to 
	  determine the scattering coefficients at the interface of the two regions. Using the asymptotic solutions at large arguments, we explicitly determine
	   the radial component of reflected current density and 
	  the scattering efficiency.
	  It is found  that the presence of a mass term outside in addition to another one inside the quantum dot strongly affects 
	the scattering of electrons. In particular, a non-null  square modulus of the scattering coefficient   is found at zero energy.
	\end{abstract}
	
\maketitle

\section{Introduction}

   Graphene  \cite{Novoselov04}  has incredible  transport properties  \cite{Stander09,Nam11} letting    it to be a potential candidate for technological applications in the future \cite{Geim07}. 
   The interaction of 
   electrons moving around the carbon atoms
   with the periodic potential of the graphene honeycomb lattice
   generates relativistic massless Dirac fermions showing a linear energy dispersion \cite{Ponomarenko8F,Katsnelson06}. These fermions have  been found to travel with a speed much faster than that of electrons in  semiconductors \cite{Semenoff84,DiVincenzo84}.
   Additionally, graphene remains capable of conducting electricity even at the limit of  nominal carrier concentration, meaning that
it never stops conducting.  In contrast,  Klein tunneling (full transmission) provides a window path for graphene electrons that limits the efficiency of electrostatic confinement \cite{Katsnelson06,Eva20}. Consequently, 
the fabrication of graphene-based materials  will remain a great challenge 
for various flexible device applications. One way to overcome such situation
 is to confine the electrons in graphene  using different techniques and then  the realization of  graphene  quantum dots (GQDs)
can offer an alternative solution.

GQDs are made up of a single atomic layer of nano-sized graphite. They  have many of the same properties as graphene, including a large surface area, a big diameter, and superior surface grafting employing $\pi-\pi$ conjugation and surface groups \cite{Shen12,Kundu19}. Recently,
 QDGs   have  been extensively discussed both theoretically \cite{Peres06, Pereira06,Silvestrov07, Matulis08, Jellal16, Jellal166, Jellal18} and experimentally \cite{Steele09,Budyka17,Liang21}. 
 Different methods have  been proposed
 to 
  confine electrons in graphene and then generate interesting systems based-graphene. These concern for instance  employing thin single-layer graphene strips 
 \cite{Peres06,Silvestrov07} or nonuniform magnetic fields \cite{Berger06}.
 Another way to 
 use  low-disorder graphene crystallographically matched to hexagonal boron nitride substrate and electrostatic confinement 
 \cite{Freitag16}. It is found that
 GQDs strongly depend on
 their sizes, shapes and nature of  edges \cite{Ponomarenko8F}. 
 GQDs can be applied  in the 
spin qubits and quantum information storage \cite{Katsnelson06,Pereira06}. Also  they can  be used  in the fields of bio-imaging, sensors, catalysis, photovoltaic devices, superconductors and so on \cite{11F}.  

  On the other hand,  systems made of gapped graphene with different  band gaps can be used to host topologically protected metallic channels in amazing ways. As a result,
   due to periodic interlayer interaction with substrates, different band gaps
   can be generated in graphene \cite{Ratnikov09,Rusponi11,Kim18}. In this context, the helical states around a mass-inverted quantum dot in graphene was studied in \cite{Myoung21}. 
   To introduce a mass-inverted quantum dot
   a heterojunction between two separate mass domains is used in similar way to the domain wall in bilayer graphene.
   It was showed that 
   the eigenstates are doubly degenerate, with each state propagating in opposing directions, preserving graphene's time-reversal symmetry.

   \par Motivated by the results mentioned above and especially \cite{Myoung21}, we study the scattering of electrons in circular graphene quantum dot
   with  mass-inverted terms through an electrostatic potential. We analytically determine 
   the solutions of energy spectrum by solving 
    Dirac equation. In some limit, we explicitly obtain the radial component of density current associated to reflected wave. This is used to compute the corresponding scattering efficiency in terms of the gaps outside and inside the quantum dot.
    The main characteristics of these quantities are studied in relation to the physical parameters of our system. 
   For this, we  identify different scattering regimes as a function of the radius, applied potential, two gaps  and  incident energy. As results, we show that the energy gap outside the quantum dot strongly affects
   the scattering of electrons in graphene.

   \par The present paper is organized as follows. In section \ref{model}, we set a theoretical model  describing our system made of circular graphene quantum dot with two gaps. 
   Subsequently, we establish the spinor solutions of the Dirac equation for the two regions. We use the continuity of the wave functions at the  boundary to explicitly determine physical quantities 
   in section \ref{scattering}. We numerically analyze 
   and discuss the main results under various  conditions  in section \ref{results}. Finally, we conclude our work.

\section{Theoretical model}\label{model}
\begin{figure}[h]
 	\centering
 \includegraphics[width=0.35\linewidth]{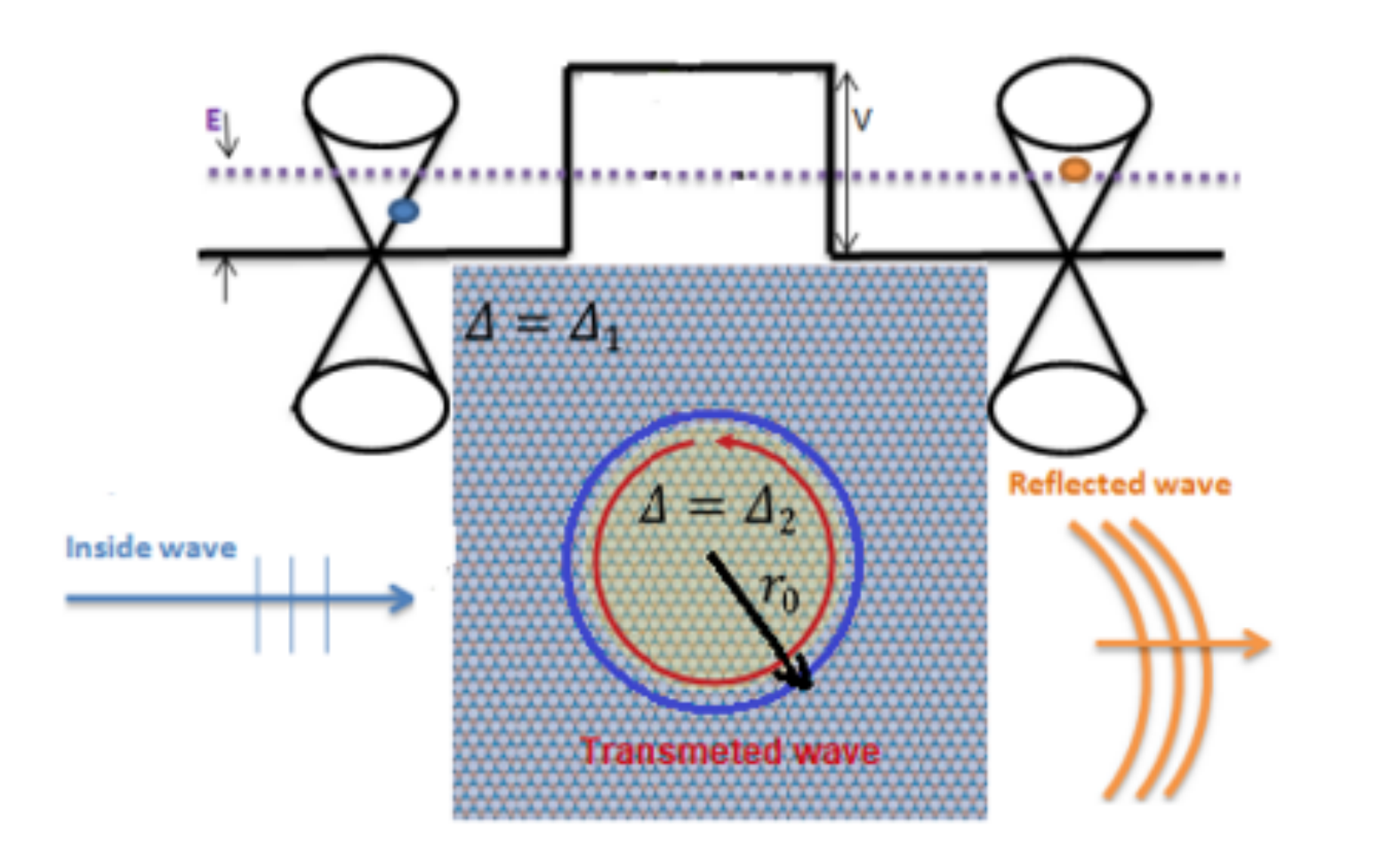}
 	\caption{(color online)  Dirac electrons propagating in a gapped graphene subject to the potential barrier
 		in a circular quantum dot of radius $ r_0 $.
 		Here the incident and reflected electron
 		waves reside in the conduction band with gap $ \Delta_{1}$, while the transmitted wave inside the dot corresponds to a state in
 		the valence band with gap $ \Delta_{2} $. }
 	\label{steady_state}
 \end{figure}

To achieve our goal, let us  consider graphene quantum dot of radius $r_0$ with
two mass terms $ \Delta(r) $ and electrostatic potential $ V(r) $. Mathematically, we have
 \begin{equation}
 	\Delta(r)=
 \left\{%
 \begin{array}{ll}
 	\Delta_1	, &  r>r_0  \\
 	\Delta_2 , &  r \leq r_0 
 \end{array}
 \right.,	
 \qquad
 V(r)=
 \left\{%
 \begin{array}{ll}
 	0, & r>r_0 \\
 	V ,&  r\leq r_0  
 \end{array}
 \right.	
 \end{equation}
 and schematically we represent our system  in Figure \ref{steady_state}.

 The  single-valley Hamiltonian describing circular graphene quantum dot can be written as  (in the  system unit $\hbar=\nu_{F}=1$)
 \begin{equation} \label{ham1}
 	H=-i\vec{\nabla}\cdot \vec{\sigma}+V(r) \mathbb{I}_2+\Delta{(r)} \sigma_z
 \end{equation}
where $\vec{\sigma}=(\sigma_x,\sigma_y,\sigma_z)$ are the Pauli matrices and $ \mathbb{I}_2$  is the  unit matrix. 
In polar coordinates $ (r,\theta)$ and   
by introducing the potentials and operators 
\begin{align}
	&
	W_{\pm}=V\pm\Delta_1\pm\Delta_2\\	
	&
	\partial_\pm= e^{\mp i\theta}
	\left(-i 
	{\partial_r}\pm\dfrac{1}{r} 
	{\partial\theta}\right)
\end{align}
we map \eqref{ham1} as
\begin{equation} \label{ham2}	
	H= \begin{pmatrix} W_{+} &  \partial_+ \\  
	\partial_-	 & W_{-} 
	\end{pmatrix}.
 \end{equation}
One can show  that the commutation relation $ [H, J_z] $ is fulfilled
by the total momentum operator  $J_{z}=L_{z}+\frac{1}{2}\sigma_{z}$ and \eqref{ham2}. This provides the separability of the eigenspinors $\psi_{m}(r,\theta)$ of the Hamiltonian \eqref{ham2} into radial $R^{\pm}(r) $ and angular $\chi^{\pm}(\theta)$ parts
\begin{equation}   
	\psi_m{(r,\theta)}=
	\begin{pmatrix} R^{+}_{m}(r)\chi^{+}(\theta)  \\R^{-}_{m+1}(r)\chi^{-}(\theta) \label{55}
	\end{pmatrix} 
\end{equation}
where the eigenstates of $J_z$ are
\begin{equation}
\chi^{+}(\theta)=\dfrac{e^{im\theta}}{\sqrt{2\pi}}\begin{pmatrix} 1\\  0  \end{pmatrix}, \qquad 
 \chi^{-}(\theta)=\dfrac{e^{i(m	+1)\theta}}{\sqrt{2\pi}}\begin{pmatrix} 0\\  1  \end{pmatrix}
\end{equation}
and $ m=0,\pm1,\pm2, \cdots $ being the angular momentum quantum number.

In order to get the solutions of the energy spectrum, we complete the derivation of the eigenspinors by  determining the radial parts. It can be achieved   by solving 
  $H\psi_m{(r,\theta)}=E\psi_m{(r,\theta)}$ in  outside $ r>r_0 $ and inside $ r\leq r_0$ regions of the quantum dot, see   Figure \ref{steady_state}.
 Indeed, for  $ r>r_0 $, we show that the radial components $ R^{+}_{m}(r)$ and $ R^{-}_{m+1}(r)$ satisfy  two coupled  differential equations
 \begin{align}
& \left(-i{\partial_r} +i\frac{m}{r}\right) R^{+}_{m}=(E+\Delta_1) R^{-}_{m+1} \label{77}
 \\
 &
	\left(-i{\partial_r} -i\frac{m+1}{r}\right) R^{-}_{m+1}=(E-\Delta_1) R^{+}_{m} \label{88}.
\end{align}
By injecting \eqref{77} into \eqref{88} we find 
a second differential equation
\begin{equation}
	\left(r^{2}{\partial_r^{2}}+ r{\partial_r}+r^{2}k_1^{2}-m^{2}\right) R^{+}_{m}(r)=0
\end{equation}
from which we find 
 the Bessel functions $ J_m(k_1r) $ as solution and set 
 the parameter
\begin{align}
k_1=\sqrt{{E^{2}}-\Delta_1^{2}}.	
\end{align}
It is convenient to  write
the  incident plane wave as
\begin{equation}
	\psi_i{(r,\theta)}
	=\dfrac{1}{\sqrt{2}}\sum_m{i^{m}}J_m(k_1r)e^{im\theta}\begin{pmatrix} 1\\  1 \label{1010}\end{pmatrix}.
\end{equation}
With this, 
 we end up with
the incident and reflected spinors 
\begin{align}
\label{inc}		\psi_i{(r,\theta)}=&\sqrt{\pi}\sum_{m}{i^{m+1}}	\left[-iJ_m(k_1r) \chi^+(\theta) +\mu_1 J_{m+1}(k_1r)
\chi^-(\theta)	\right]\\
\label{ref}
	\psi_r{(r,\theta)}=&\sqrt{\pi}\sum_{m}{i^{m+1}}a_{m}	\left[-iH^{(1)}_m(k_1r)\chi^+(\theta)+  \mu_1H^{(1)}_{m+1}(k_1r)
\chi^-(\theta)	\right]	
\end{align}	
  where $ H^{(1)}_{m}(k_1r) $ is the Hankel function of the first kind, $ a_{m} $ are the scattering coefficients. Here 
  we have defined dimensionless parameter
  \begin{align}
  	\mu_1=\sqrt{\frac{E-\Delta_1}{E+\Delta_1}}
  \end{align}
 in terms of the first energy gap $\Delta_{1}$ and it reduces to one for 
  $\Delta_{1}=0$.

 Now we consider the second region $ r\leq r_0$ with 
the potential $V$ and energy gap $\Delta_{2}$. As a result, we  find the following equations 
 \begin{align}
	& \left(-i{\partial_r} +i\frac{m}{r}\right) R^{+}_{m}=(E-V+\Delta_2) R^{-}_{m+1} \label{1313}
	\\
	&
\left(	-i{\partial_r} -i\frac{m+1}{r}\right) R^{-}_{m+1}=(E-V-\Delta_2) R^{+}_{m}\label{1414}
\end{align}
giving rise to 
\begin{align}\label{trss}
	(r^{2}{\partial_r^{2}}+ r{\partial_r}+r^{2}k_2^{2}-m^{2}) R^{+}_{m}=0
\end{align}
and we have set 
\begin{align}
	k_2=\sqrt{(E-V)^{2}-\Delta_2^{2}}.
\end{align}
From \eqref{trss} we derive the transmitted spinor solution
\begin{align}
	\psi_t{(r,\theta)}=\sqrt{\pi}\sum_{m}{i^{m+1}}b_{m}	\left[-iJ_m(k_2 r)\chi^+(\theta)
+\mu_2 J_{m+1}(k_2 r) \chi^-(\theta)
\right]\nonumber
\end{align}
where $ b_{m }$ are the scattering coefficients
and
\begin{align}
 \mu_2 =\sqrt{\frac{E-V-\Delta_2}{E-V+\Delta_2}}.
\end{align}
In the next, we will see how the above results can be used to study the scattering  of Dirac electrons in our system with
the presence of  two mass terms.

\section{Scattering problem}\label{scattering}

To  study the scattering problem associated our system, in the first stage we     determine 
the scattering coefficients $ a_{m}$ and $ b_{m}$. To this end, we
 use of the boundary condition at interface  
$ r=r_0 $ to write
\begin{align}
 \psi_i{(r_0)}+\psi_r{(r_0)}=\psi_t{(r_0)}. 
 \end{align}
After substitution, we establish two relations between Bessel and Hankel functions
\begin{align}
&	J_{m}(k_1r_0)+a_{m}H^{(1)}_{m}	(k_1r_0)=b_{m}J_{m}	(k_2 r_0)	
\\
&
	\mu_1J_{m+1}(k_1r_0)+\mu_1a_{m }H^{(1)}_{m+1}	(k_1r_0)= \mu_2 b_{m}J_{m+1}	(k_2 r_0)			
\end{align}
which
 can be solved to obtain the scattering coefficients
\begin{align}
&a_{m}=\frac{{ \mu_2J_{m}(k_1r_0)J_{m+1}	(k_2 r_0) -\mu_1J_{m}(k_2 r_0)J_{m+1}(k_1r_0)}}{\mu_1 J_{m}(k_1 r_0)H^{(1)}_{m+1}	(k_1r_0)-\mu_2 J_{m+1}	(k_2 r_0)H^{(1)}_{m}	(k_1r_0)}\label{amm}				
\\
&
 b_{m}=\frac{{\mu_1J_{m}(k_1 r_0)H^{(1)}_{m+1}(k_1r_0)- \mu_1J_{m+1}	(k_1 r_0)H^{(1)}_{m}(k_1r_0) }}{\mu_1J_{m}	(k_1r_0)H^{(1)}_{m+1}	(k_1r_0)-\mu_2 J_{m+1}	(k_2 r_0)H^{(1)}_{m}	(k_1r_0)}.	
\end{align} 

At this level, we compute the radial component of current density  corresponding to our system. For this, we use the Hamiltonian \eqref{steady_state} to obtain the current density 
\begin{equation}
\vec{j}=\psi^{\dagger}\vec{\sigma}\psi  
\end{equation}
 where inside  the quantum dot $\psi=\psi_{t}$ and outside  $ \psi=\psi_{i}+\psi_{r}$. From the projection
 \begin{align}
 j_{r}=\vec{j} \cdot \vec{e}_r	
 \end{align}
 one obtains
\begin{equation} j_r=\psi^{\dagger} \begin{pmatrix} 0& \cos{\theta}-i\sin{\theta} \\ \cos{\theta}+i\sin{\theta}&0 \end{pmatrix}\psi.
\end{equation} 
As far as
the reflected wave \eqref{ref}  is concerned, we get 
\begin{align} \label{jrr}
	j^{r}_r=\dfrac{1}{2}\sum_{m=0}^{\infty} A_{m}(k_1r)\begin{pmatrix} 0& {e^{-i\theta}} \\e^{i\theta}& 0 \end{pmatrix}\sum_{m=0}^{\infty} B_{m}(k_1r)
\end{align} 
such that
\begin{align} \label{311}
&A_{m} (k_1r)
=(-i)^{m+1}\left[iH^{(1)^{*}}_{m}(k_1r)
\begin{pmatrix}a^{*}_{m}e^{im\theta}& a^{*}_{-m-1}  e^{-im\theta}\end{pmatrix}+\mu_1H^{(1)}_{m+1}(k_1r)\begin{pmatrix}a^{*}_{-m-1}e^{i(m+1)\theta}& a^{*}_{m}e^{-i(m+1)\theta}\end{pmatrix}\right]
\\
& B_{m} (k_1r)
=(-i)^{m+1}\left[iH^{(1)}_{m}(k_1r)\begin{pmatrix}a^{*}_{m}e^{-im\theta}\\a^{*}_{-m-1}e^{im\theta}\end{pmatrix}+\mu_1H^{(1)^{*}}_{m+1}(k_1r)\begin{pmatrix}a^{*}_{-(m+1)}e^{i(m+1)\theta}\\a^{*}_{m}
	e^{-i(m+1)\theta}\end{pmatrix}\right].\label{322}
\end{align} 

To illustrate our results and give a better understanding let us consider 
the asymptotic behavior of the Hankel function  for large argument $k_1r\gg1$. With this we will be able to explicitly establish analytical results of the above quantities. Then in such limit, one can use
the approximate function 
\begin{equation} H_{m}(k_1r)\simeq\sqrt{\dfrac{2}{\pi k_1r}}e^{(k_1r-\frac{m\pi}{2}-\frac{\pi}{4})}\label{2626}
\end{equation}
which can 
 be injected into (\ref{311}-\ref{322})   
to approximate the radial component \eqref{jrr}
by
\begin{equation}
j^{r}_r=\frac{2}{\pi k_1 r}\sum_{m=0}^{\infty}|c_m|^{2} \left(\cos[(2m+1)\theta](1+\mu_1^{2})+2\mu_1\right)
\label{jrra}
\end{equation} 
where the square modulus of the scattering coefficients
\begin{align}\label{cmm}
|c_m|^{2}=\frac{1}{2}(|a_m|^{2}+|a_{-(m+1)}|^{2})	
\end{align}
are given in terms of  $a_m$ \eqref{amm}.
Actually
(\ref{jrra}-\ref{cmm}) show a strong dependence on the energy gap $\Delta_{1}$ outside the quantum dot that
is not the case for analogue results obtained in \cite{Jellal18}. As a consequence,  we will study the  rule will be played by   $\Delta_{1}$ 
to affect the scattering problem of our system.

Let us analyze other physical quantities related to the radial current for the reflected wave \eqref{jrra}. Indeed, 
the scattering cross-section $\sigma $ is defined by
\begin{equation}
\sigma=\frac{I^{r}_r}{(I_i/A_u)}	
\end{equation}
such that   the total reflected flux per unit area $I^{r}_r $ can be calculated 
by integrating \eqref{jrra}
\begin{align}
	I^{r}_r=\int^{2\pi}_{0}j^{r}_r(\theta)r d\theta
\end{align}
and as a result we find
\begin{equation} 
 I^{r}_r	
	=\frac{8}{E+ \Delta_{1}}\sum_{m=0}^{\infty}|c_m|^{2}\label{3030} \end{equation}
while  for the incident wave \eqref{1010}  the ratio $I_i/A_u$ is just the unit
 and therefore the cross-section reduces to $I^{r}_r $, i.e.
 $ \sigma= I^{r}_r$.

To go deeply in the study of the scattering problem  for  Dirac electron in graphene circular  quantum dot, we consider the scattering efficiency $ Q $. It  is defined by the ratio between the scattering  cross-section  and the geometric cross-section
\begin{align}
Q=\frac{\sigma}{2r_0}	
\end{align}
and then using \eqref{3030} to get
  \begin{equation}
	Q=\frac{4}{r_0(E+\Delta_{1})}\sum_{m=0}^{\infty}|c_m|^{2}\label{3031}.
\end{equation}
We emphasis that in the case $\Delta_{1}=0$ our results reduce to 
those obtained in  \cite{Jellal18}.
We will numerically show how the presence of
$\Delta_{1}$ will affect the scattering problem.

\section{Results and discussions}  \label{results}

\begin{figure}[H]	
	\centering
	{\includegraphics[scale=0.66]{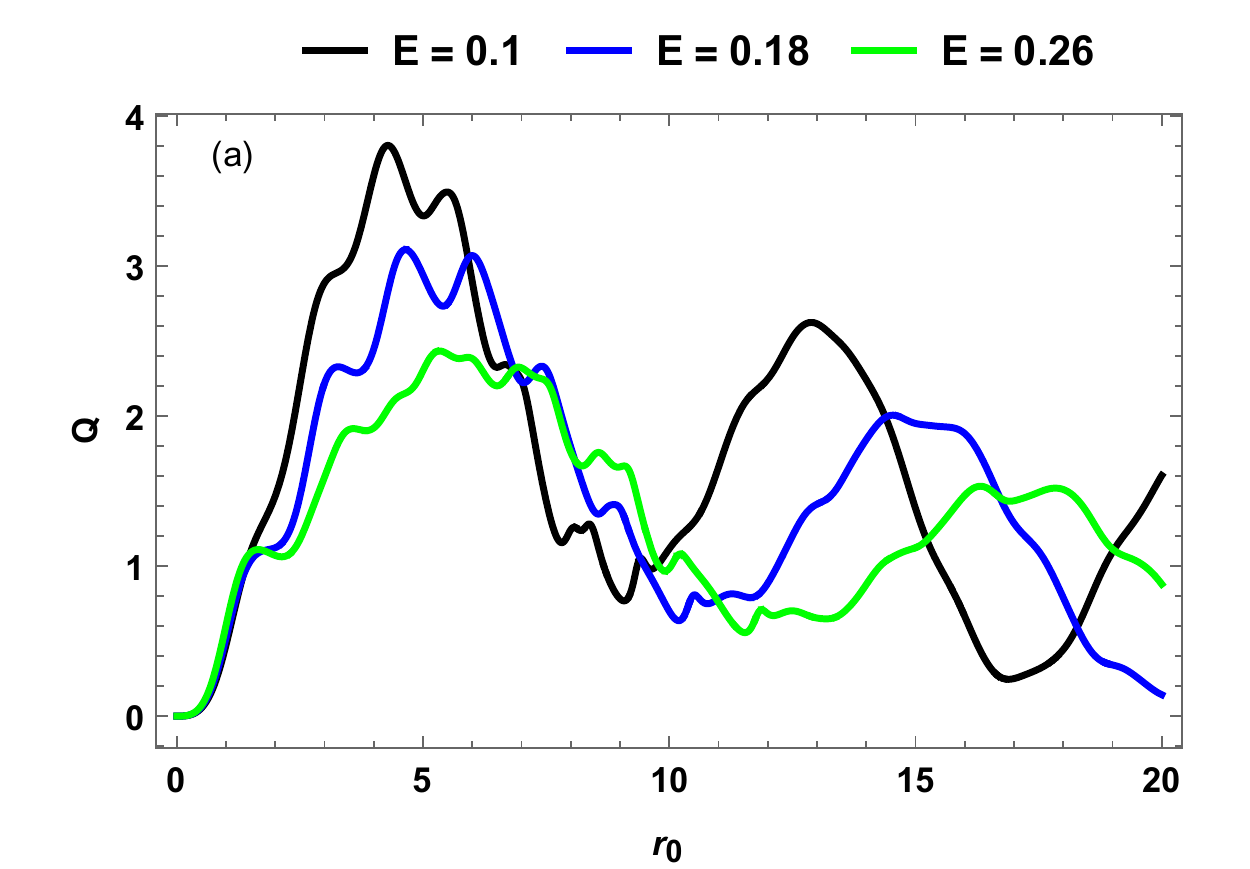}}	
	{\includegraphics[scale=0.65]{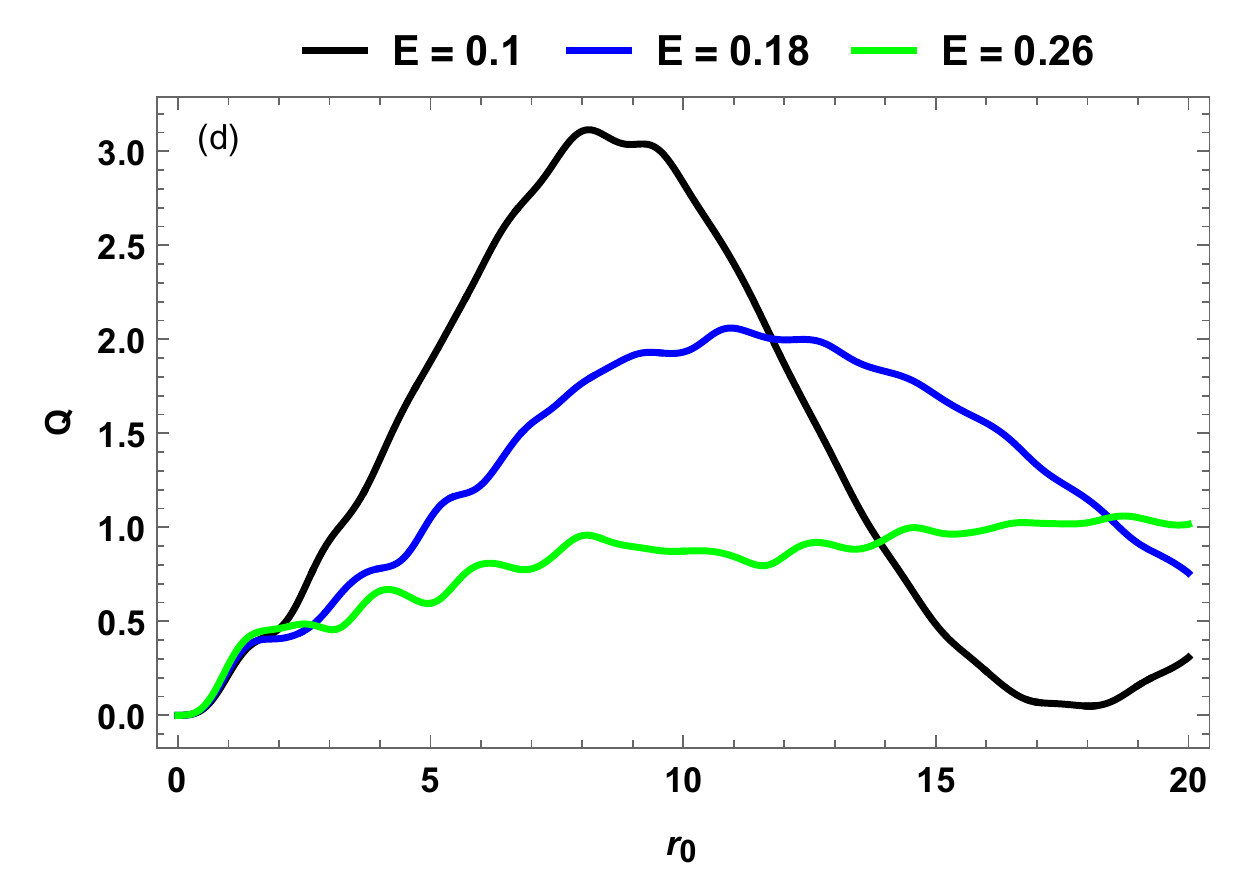}}\\
	{\includegraphics[scale=0.65]{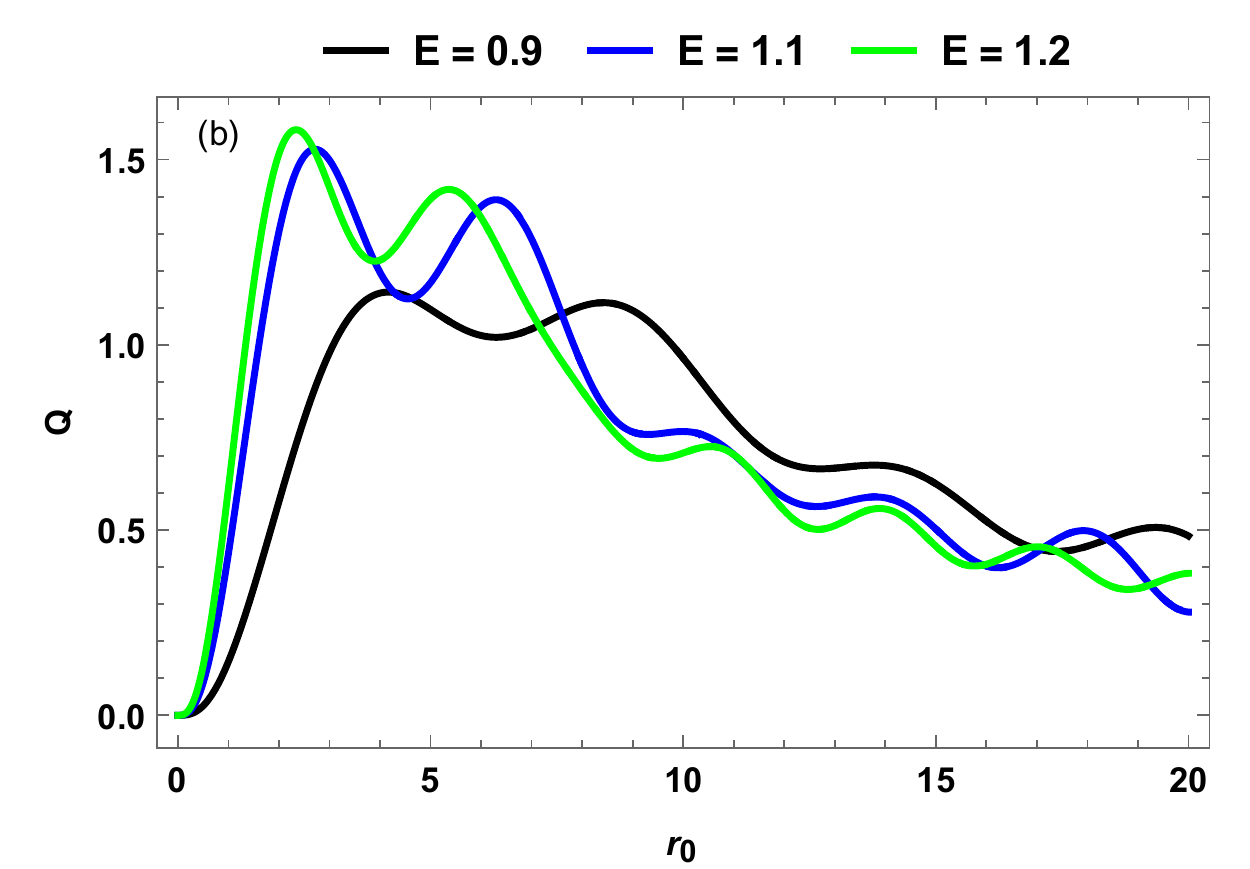}}
	{\includegraphics[scale=0.65]{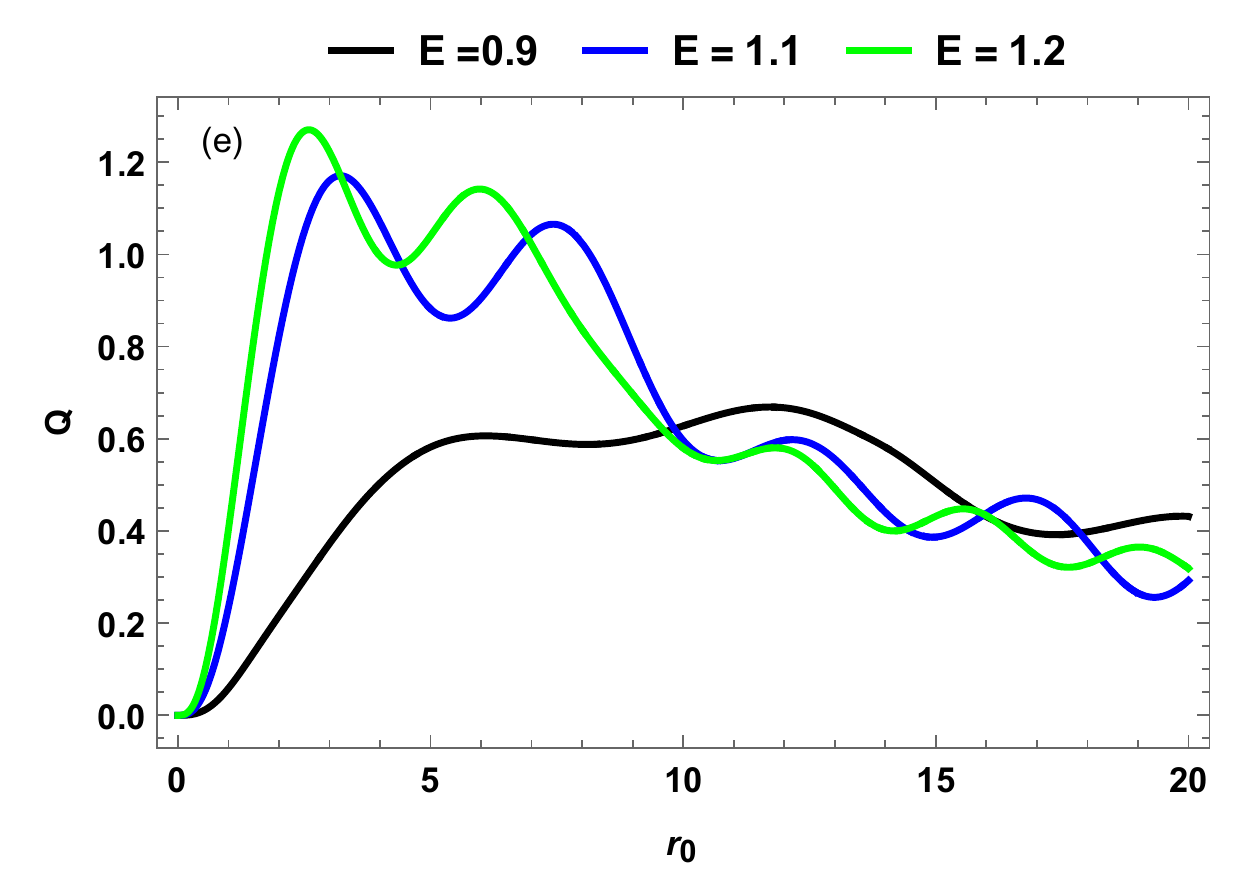}}	\\
	{\includegraphics[scale=0.65]{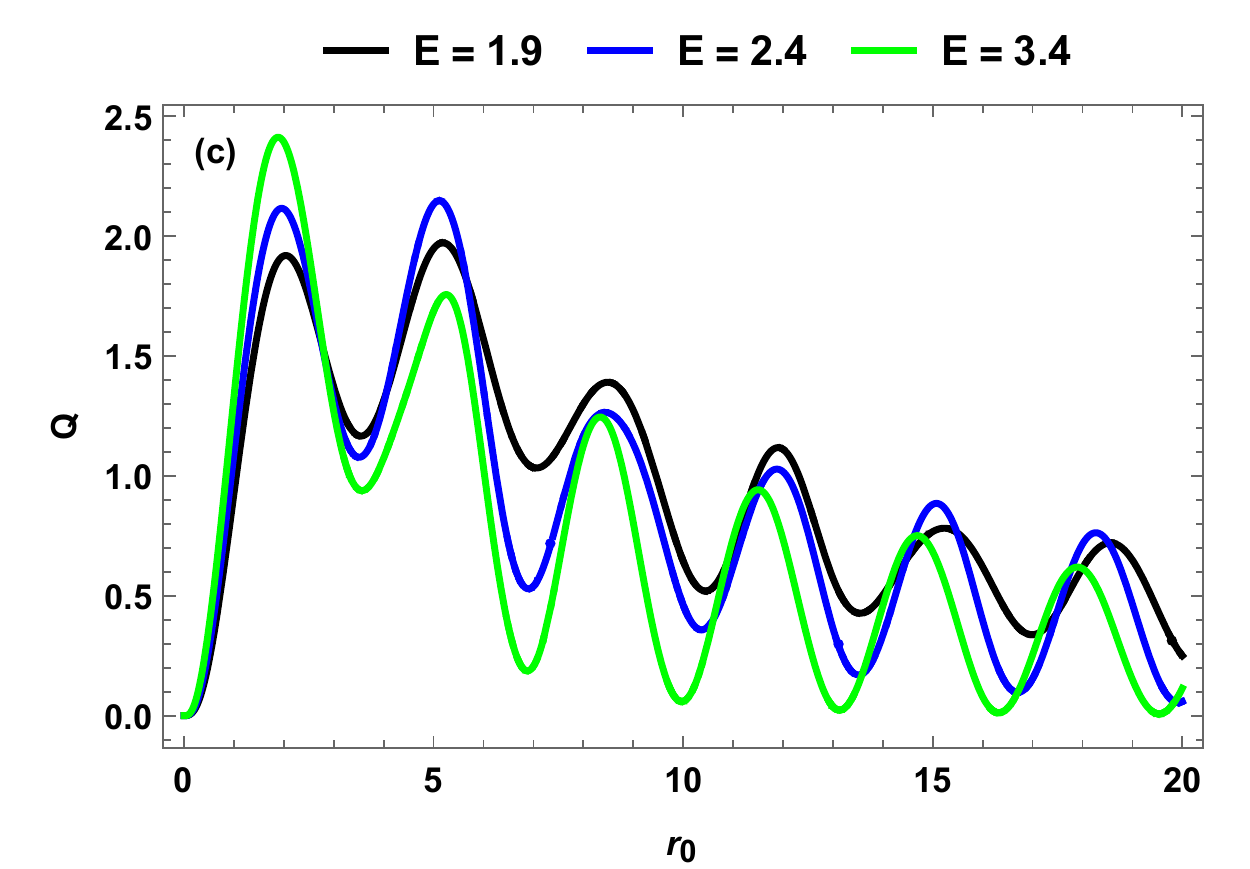}}
	{\includegraphics[scale=0.65]{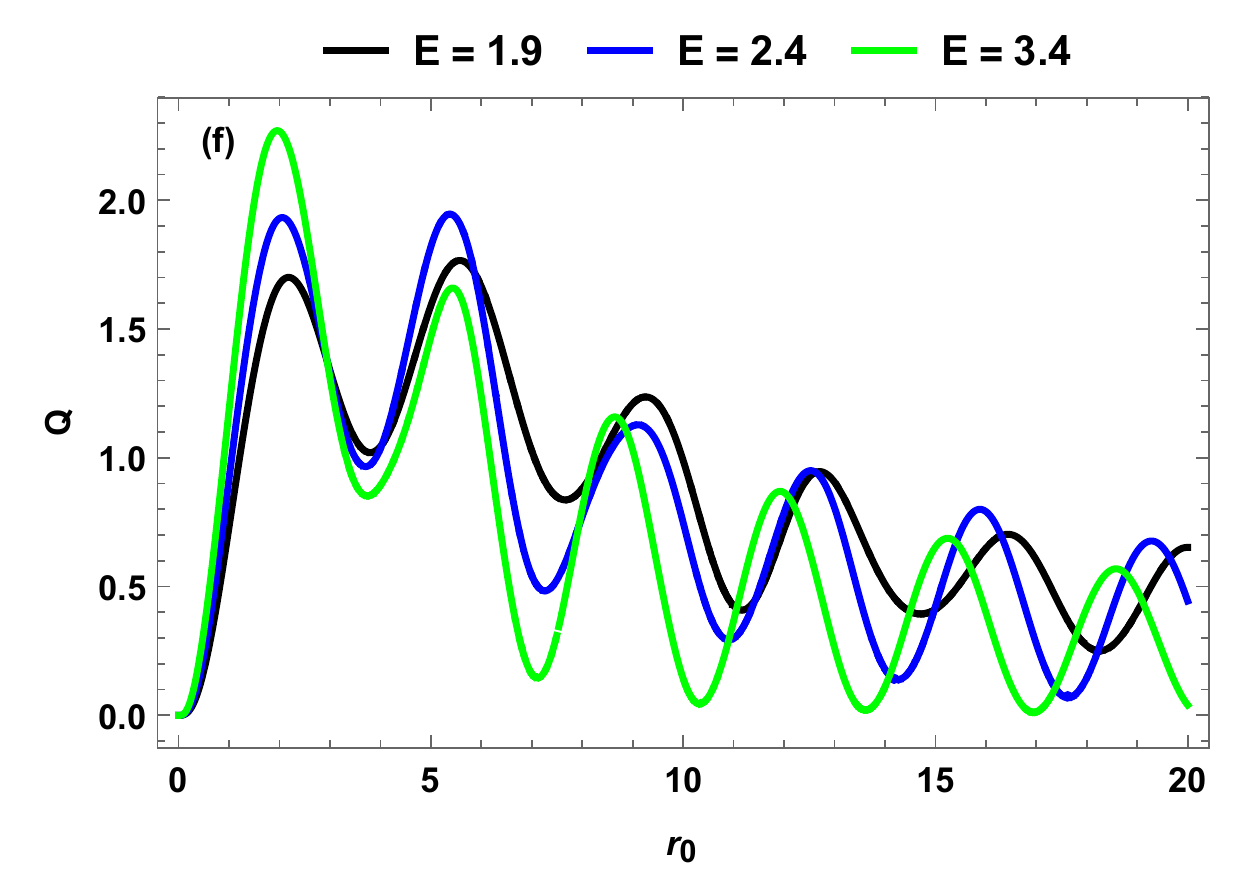}}\\
	\caption {{(color online) {The scattering efficiency $ {Q} $ versus the quantum dot radius $r_0$ for $ \Delta_2=0.2$ and   $V=1$. Three
		regimes of the incident  energy are considered  (a,d): $ E<W_-$, (b,e): $ W_-< E<W_+$, (c,f): $ E> W_+$}. Here (a,b,c): $\Delta_1=0.5 $  and    (d,e,f):   $\Delta_1=0.7 $.}}\label{a}
	
\end{figure}

We proceed a numerical analysis by considering various conditions
 of the physical parameters to underline the main features of our system. Indeed,  
Figure \ref{a} represents the scattering efficiency $ Q $ as a function of the   quantum dot radius $r_0$. Here we choose 
  $ V=1 $ and $\Delta_2=0.2$, with 
 (a,b,c):  
  $\Delta_1=0.5$ and  (d,e,f): $\Delta_{1}=0.7$. 
Three scattering  regimes of the incident energy   $E$ are considered.
 Figures \ref{a}(a,d) show the results for the regime $ E<W_-$   outside the quantum  dot. We notice that $Q$  increases linearly for small radius, but when the radius reaches a certain value $Q$ begins to change by showing a oscillatory behavior  until reaches a maximum. By increasing further $r_0$, we observe
 a decrease   on the oscillations shape of $Q$. We notice that $Q$ is very sensitive to the incident energy $E$ because it decreases when  $E$
 increases. In addition for $\Delta_1=0.5$ in Figure \ref{a}(a) there is a maximum value $Q=4$ when $r_0$ tends to $4$. While for $\Delta_{1}=0.7$ in Figure \ref{a}(d)  the  maximum value is $Q=3$ when $r_0$ tends towards $8$.  
 The period of oscillations corresponding to $\Delta_1=0.5$  is small compared to the case $\Delta_{1}=0.7$. 
Figures \ref{a}(b,e) show the behavior of   $Q$  for the electronic state inside the quantum dot   $ W_-< E<W_+ $. It is clearly seen  that $ Q $ increases linearly {{up to a certain value of $r_0$, then $Q$ changes in oscillatory shape with different amplitudes  depending on the incident energy $ E $. In fact, for $\Delta_{1}=0.5$ when $r_0$ tends to $6$ and for $\Delta_{1}=0.7$ when $r_0$ tends to $10$ these amplitudes changes inversely with incident energy  values, and afterword the oscillations begin to be damped.}}
The last  regime of energy $ E> W_+$ is presented in 
Figures \ref{a}(c,f). As one sees for small radius,  $Q$ changes in a linear manner  and  its values  are very close to each other regardless of the incident energy. However, when the radius reaches a certain value, $Q$ changes by oscillating with  decreasing amplitudes  until a certain value of $r_0$. These results show the influence of the gap $\Delta_{1}$
on the scattering efficiency  $Q$.

\begin{figure}[H]
	\centering
	{\includegraphics[scale=0.65]{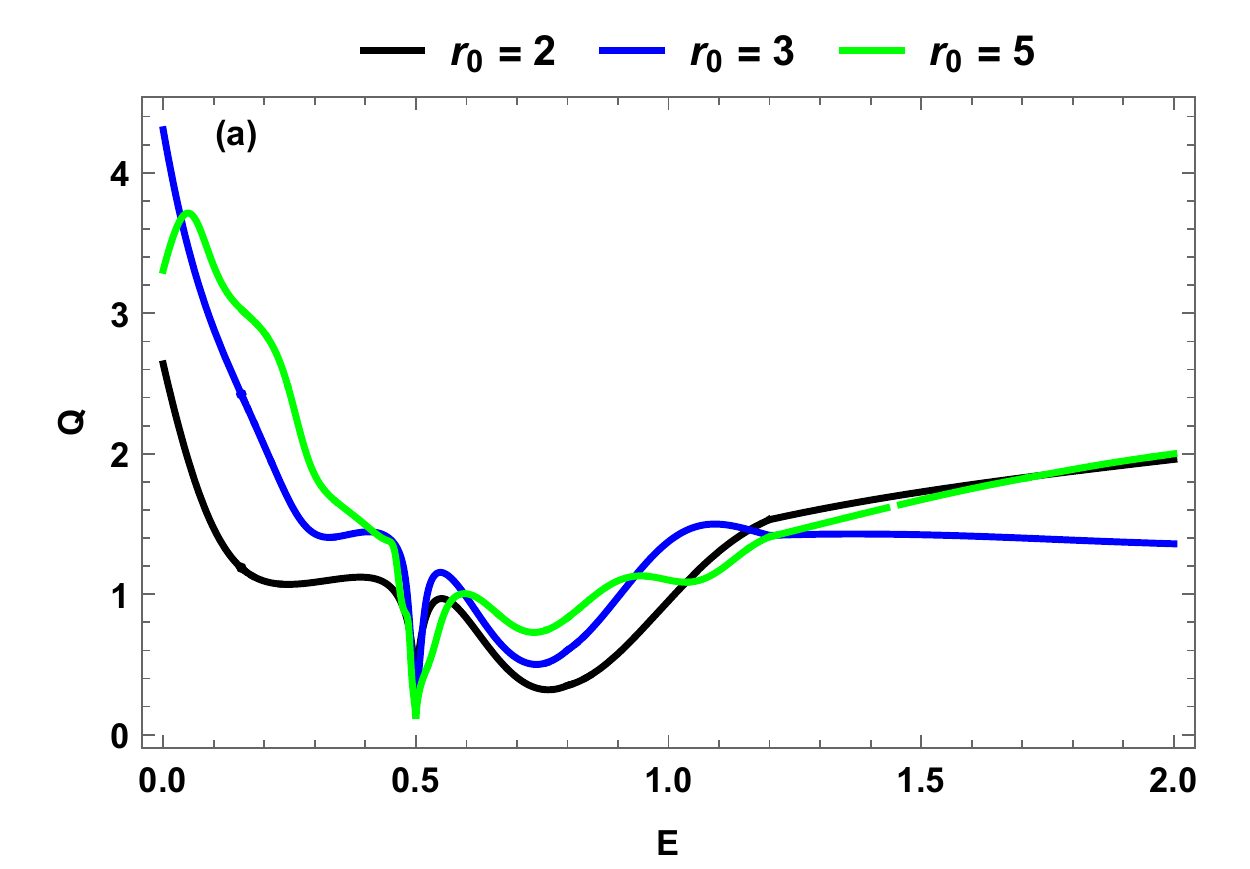}}
	{\includegraphics[scale=0.65]{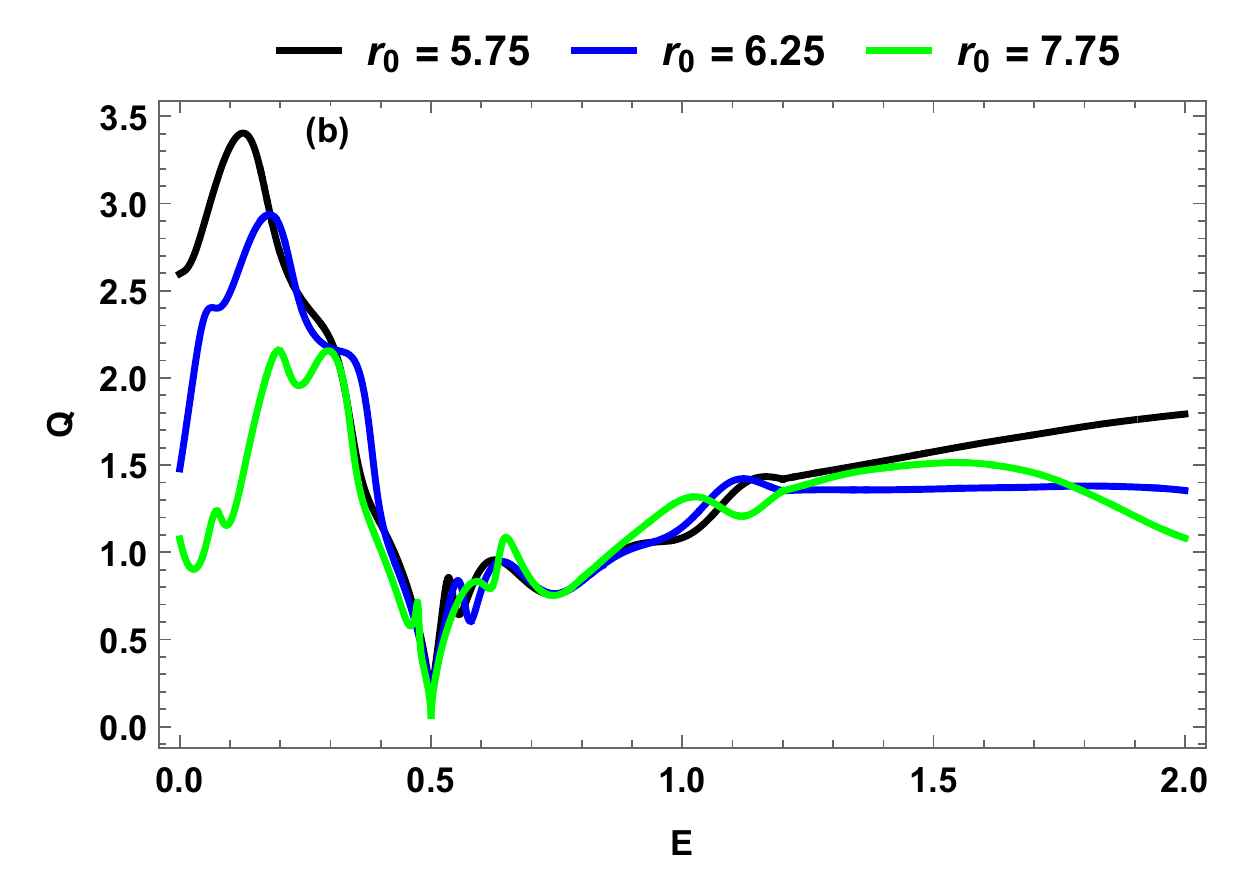}}\\
	{\includegraphics[scale=0.65]{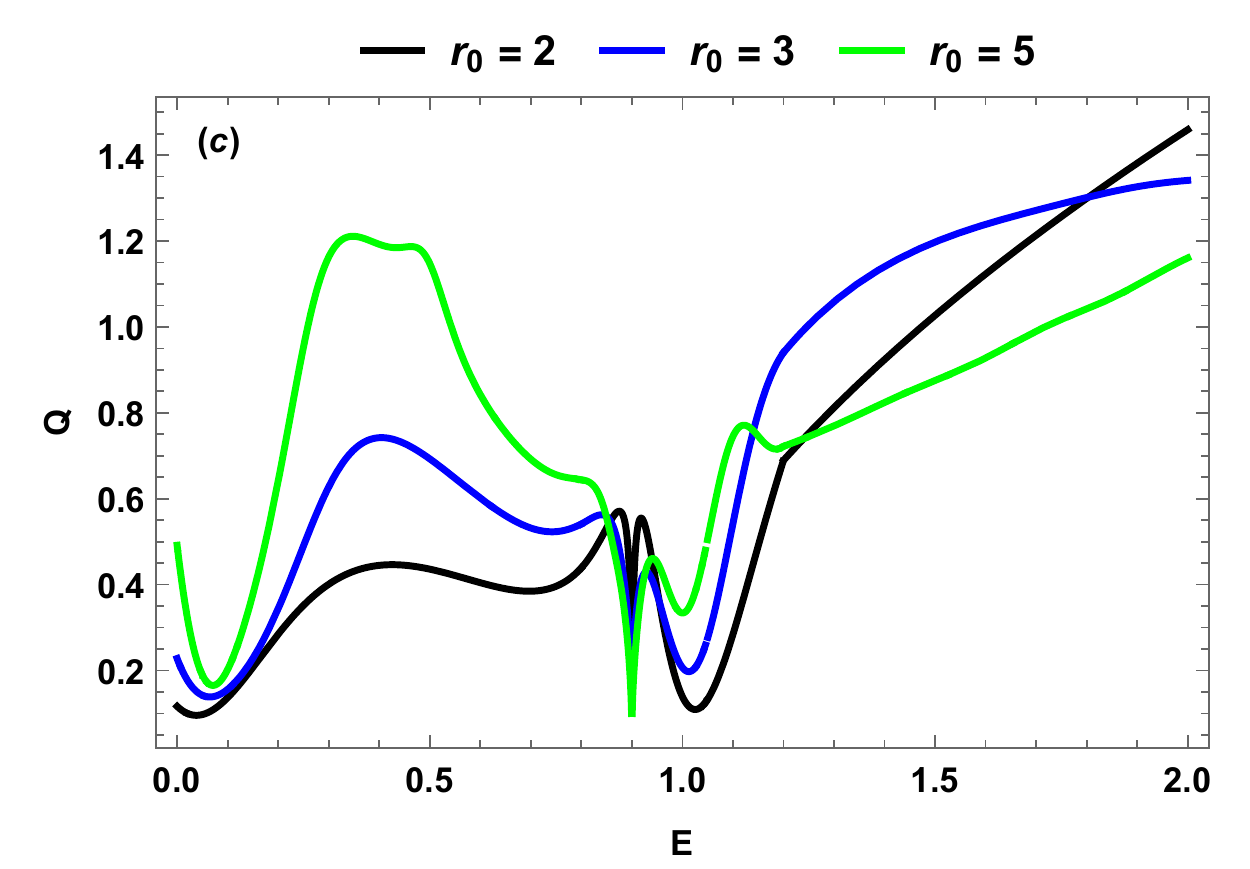}}
	{\includegraphics[scale=0.65]{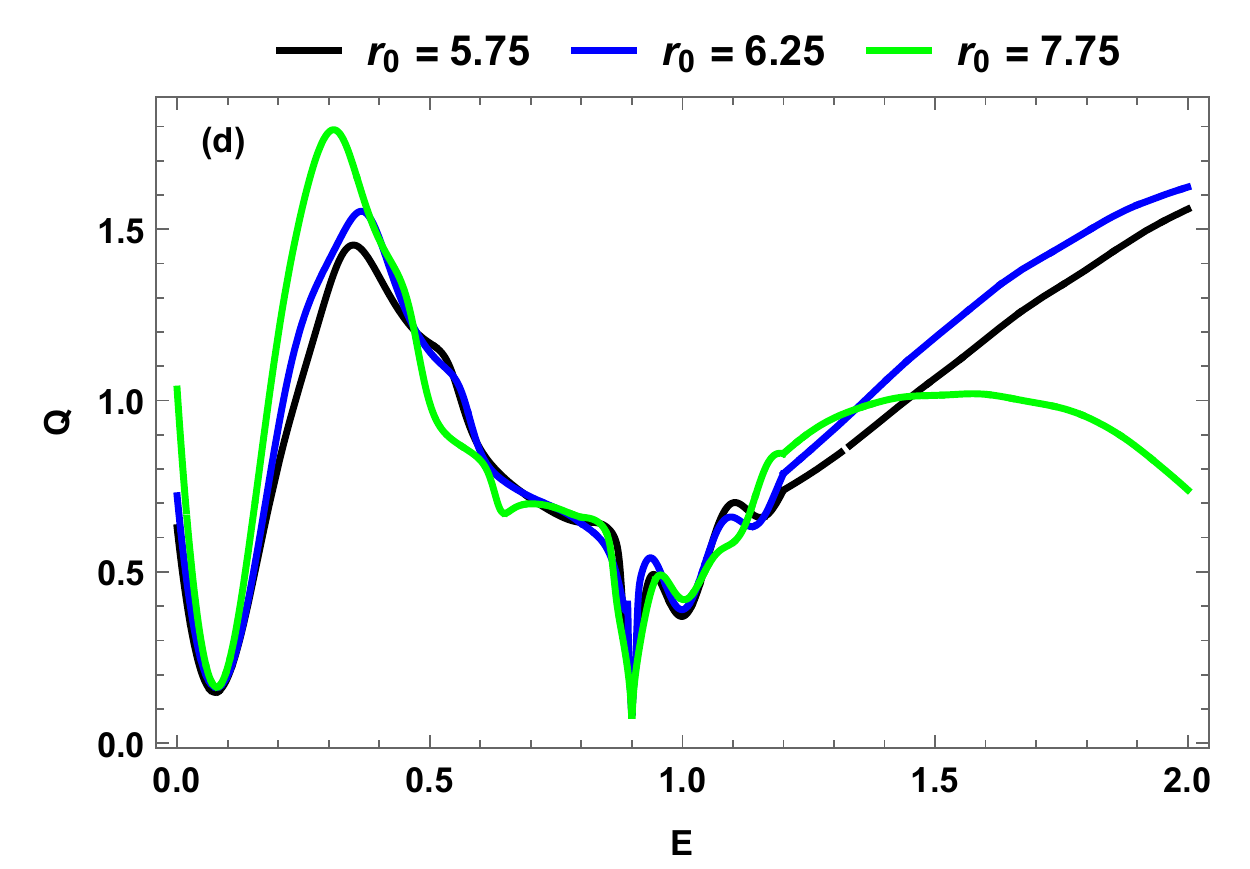}}
	\caption{{(color online) The scattering  efficiency $Q$ versus the incident  energy $E$  for 
	$	\Delta_2=0.2$ and $V=1$.
	Different sizes	
			 of the quantum dot are considered. Here (a,b):  $\Delta_1=0.5 $ and    (c,d): $\Delta_1=0.9 $.}}
	\label{f3}
\end{figure}

We plot in Figure \ref{f3} the scattering efficiency $Q$ as a function of the incident energy $E$ for $V=1$, $\Delta_{2}=0.2$, two values $\Delta_{1}=0.5,0.9$ and different sizes  of the quantum dot. For small values of $r_0$ in Figures \ref{f3}(a,c)   $ Q $ takes 4 as  maximum value  for $E= 0$ and $\Delta_1=0.5$. It decreases rapidly until it approaches zero  at  $E=\Delta_{1}=0.5$.  
When the energy increases $Q$ gradually increases by oscillating to reach constant values. For $\Delta_{1}=0.9$ we observe that $Q$ is minimal
at $E=0$ but it reaches the	maximum value $1.2$ by increasing $E$. Later on it
		 decreases to approaches zero at  $E=\Delta_{1}=0.9$ and after that both of them increase.
For large radius $r_0$ in Figures \ref{f3}(b,d) $Q$ is not zero for $E=0$ and shows large maxima. The maximum value of $Q$ decreases when $r_0$ increases then decreases towards a value close to zero for $E=\Delta_{1}=0.5$. By increasing $E$, $Q$ takes constant values and strongly depends on $r_0 $. {{As for $\Delta_{1}=0.9$,   $Q$ has a  minimum value at $E=0$ and reaches   $1.8$ as maximum
		when $ E $ increases. It
		decreases to approaches zero at  $E=\Delta_{1}=0.9$ but once  $ E $ increases  we observe an increase of $Q$  for both sizes  $r_0=5.75, 6.25$}}. 

\begin{figure}[H]
	\centering
		{\includegraphics[scale=0.66]{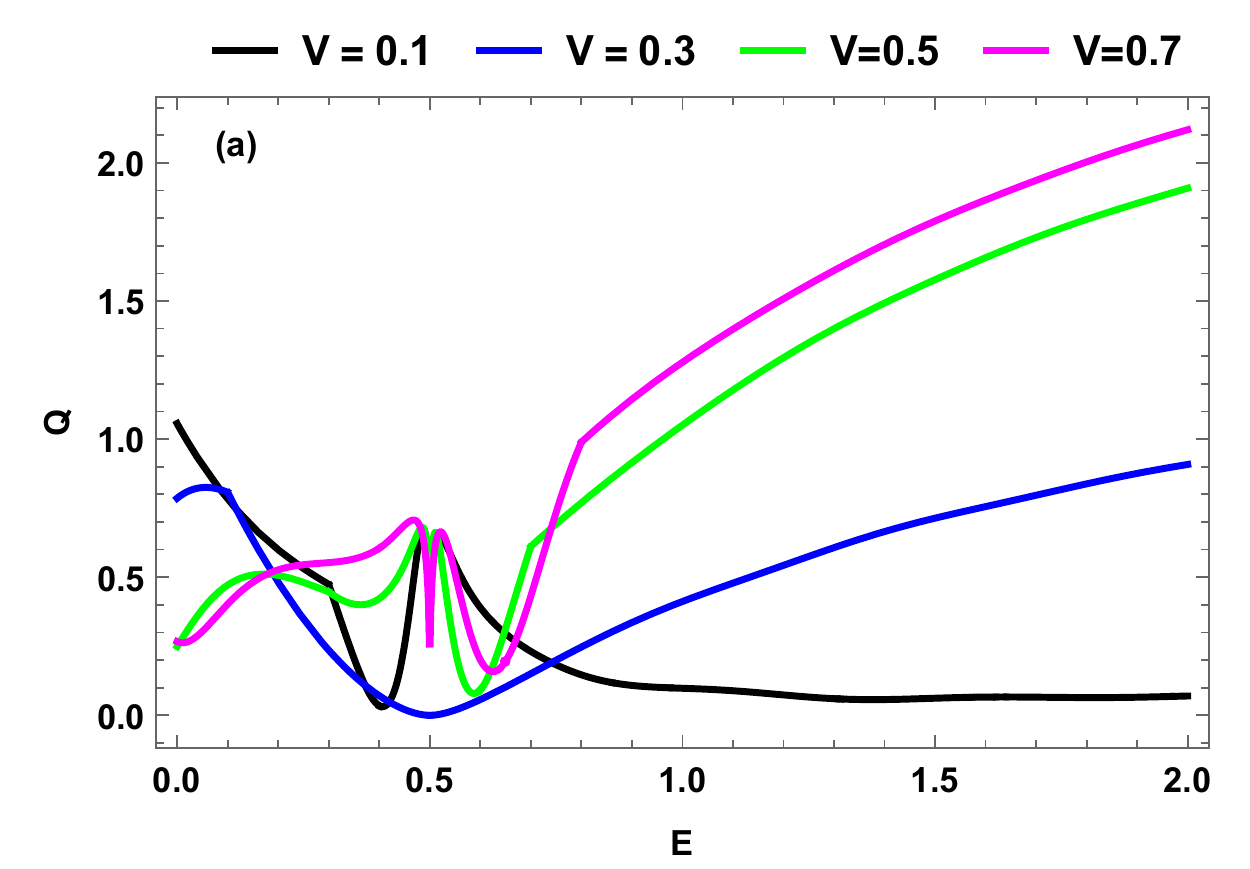}}
	{\includegraphics[scale=0.6]{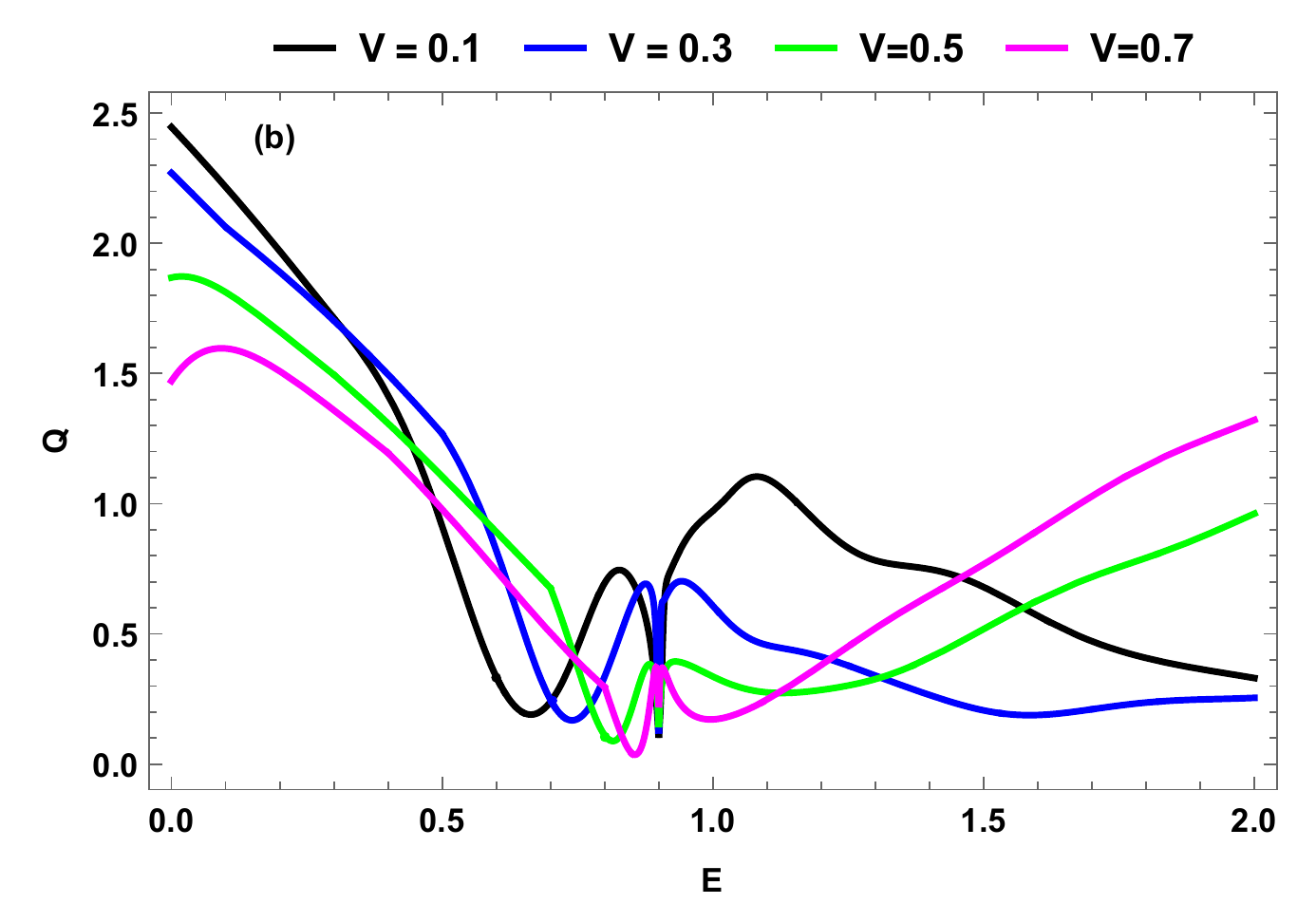}}
	\caption{{(color online) The scattering efficiency $ Q $ versus  the incident energy $ E $ for  $ \Delta_2=0.2 $ and  $r_0= 3$.
			Different values  of the potential $ V $ are considered. Here (a): $ \Delta_1=0.5 $  and  (b): $ \Delta_1=0.9 $.}} \label{f5}
\end{figure}

In order to show how the potential $V$  affects the scattering  we represent in Figure \ref{f5} the scattering efficiency $Q$ as a function of the incident energy $E$  
for		
$\Delta_2= 0.2 $ and $r_0 =3$ with  (a): $ \Delta_1=0.5$ and (b): $\Delta_{1}=0.9$. It is clearly seen that when $E$ closes to zero, $Q$ is not null and its values are strongly depending on both parameters $V$ and $ \Delta_1$. Moreover, as long as $E$ increases we observe that $Q$ decreases towards $ \Delta_1$ by showing  
 small oscillations with different amplitudes. Afterwords,  one sees that  $Q$ starts to increase  by increasing $E$ for large potential $V$. An interesting conclusion can be emphasized  
 such that for $E<\Delta_1=0.5 $ in Figure \ref{f5}, $Q$ takes small values compared to the case $E< \Delta_1=0.9$ in Figure \ref{f5}b. However,   for the case 
 $E>\Delta_1=0.5 $ and 
 $E> \Delta_1=0.9$ we observe the opposite behavior.

\begin{figure}[H]
 	\centering
 	{\includegraphics[scale=0.65]{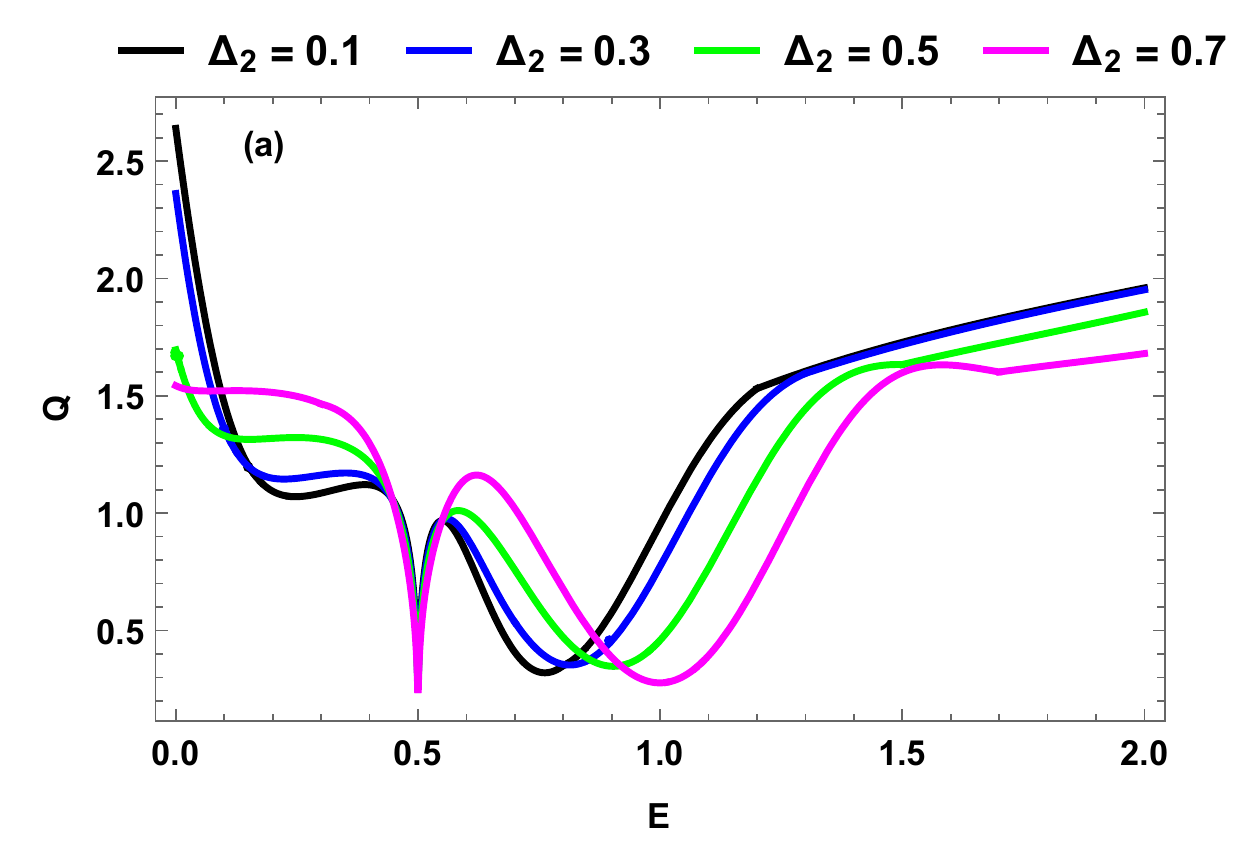}}
 	{\includegraphics[scale=0.65]{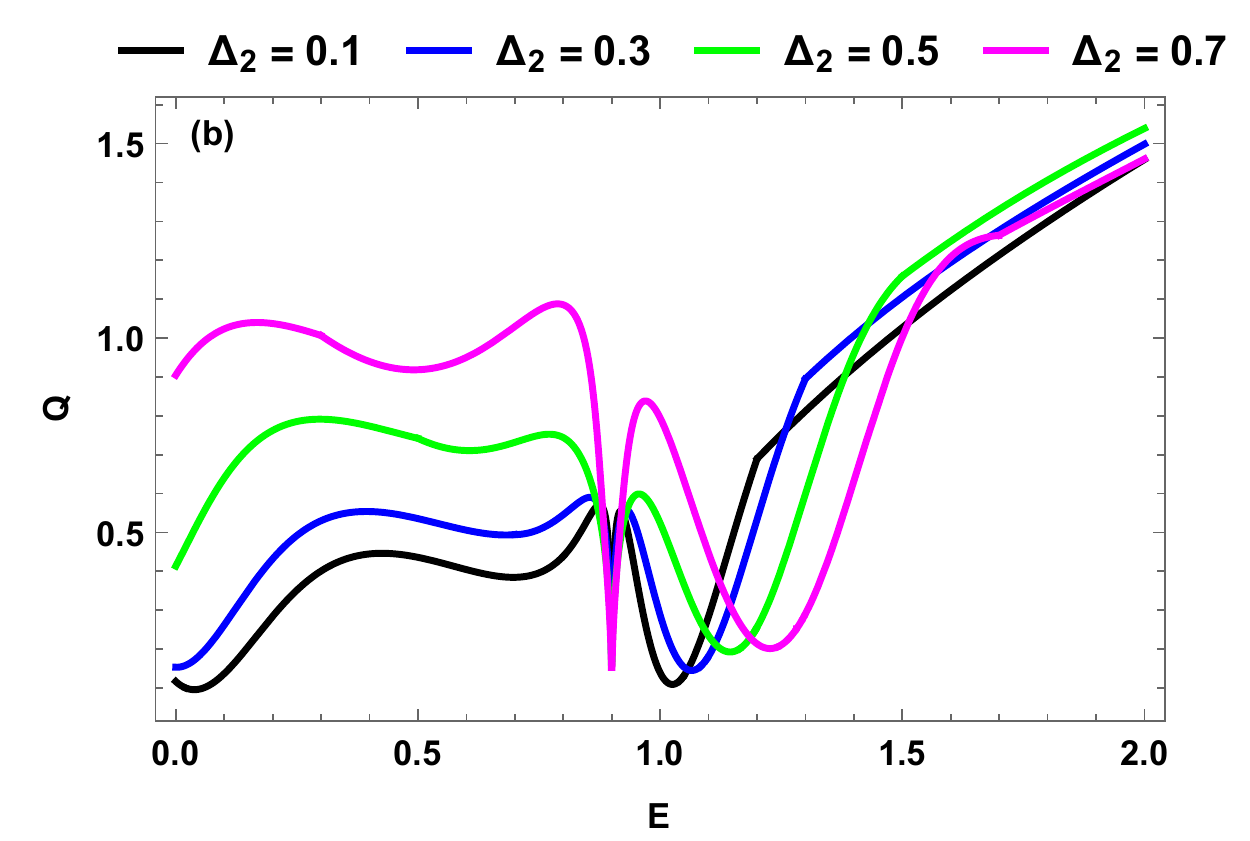}}
 	\caption{{(color online) The scattering efficiency $ Q $ versus  the incident energy $ E $  
 	for $ V=1$ and $r_0= 2$. Different values of the gap inside quantum dot $\Delta_2$ are considered. Here,  (a): $ \Delta_1=0.5$ and (b): $ \Delta_1=0.9 $.}}\label{f6}
 \end{figure}

  Figure \ref{f6} represents  the scattering efficiency $Q$ as a function of the incident energy $E$ for different values of energy gap {{ $ \Delta_2$ inside the quantum dot with $V=1$ and
  		$r_0=2$. Here we choose  two values such that (a): $\Delta_{1}=0.5$ and (b): $\Delta_{1}=0.9$. It is clearly seen that $Q $ is not null  for $E = 0$. Under the increase of $E$ we observe that $Q$ decreases by approaching zero at the point $E=\Delta_{1}=0.5$ as shown  in Figure  \ref{f6}a. Afterward, $Q$  oscillates to reach constant values depending on  $ \Delta_2$. Now  in Figure  \ref{f6}b for $\Delta_{1}=0.9$ 
  		one sees that $Q$ shows mostly the same behavior
  		regardless the value of  $ \Delta_2$ in the interval $0<E<\Delta_1=0.9$ 
  		except that it increases.    Whereas   for $E>\Delta_{1}=0.9$}}
we observe that $Q$ presents some oscillations and afterward it increase. 
 
\begin{figure}[H]
	\centering
	{\includegraphics[scale=0.65]{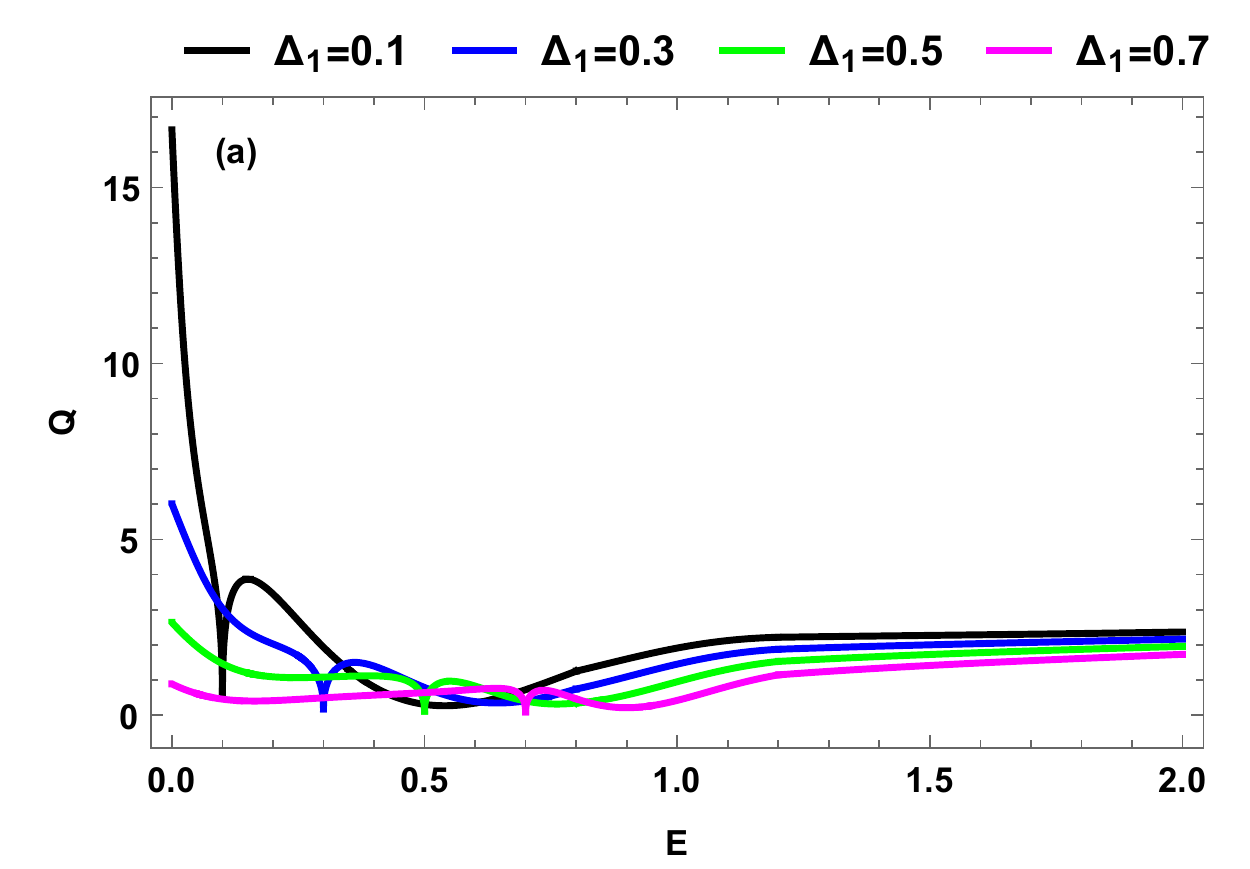}}
	{\includegraphics[scale=0.65]{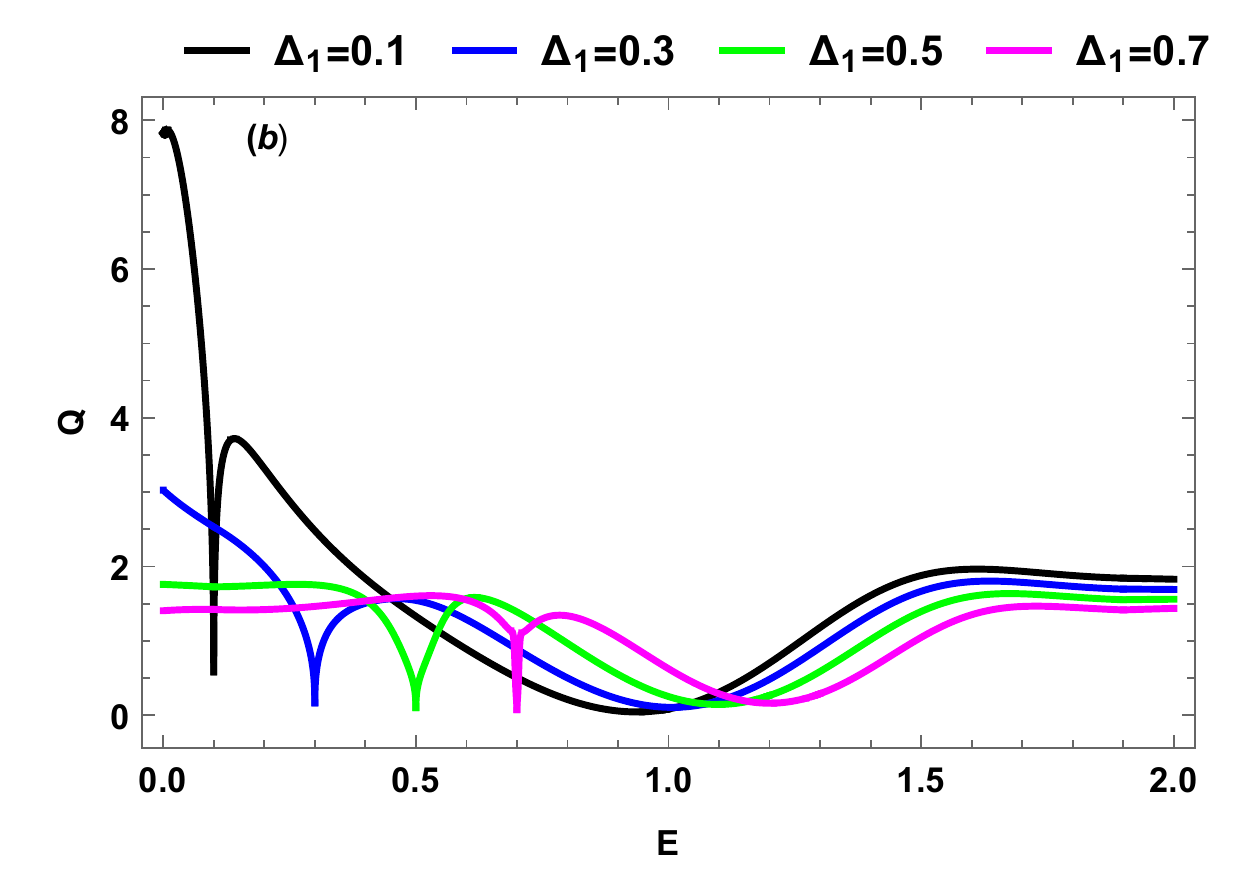}}
	\caption{{(color online) The scattering efficiency $ Q $ versus  the incident energy $ E $ for  $V=1$,   $r_0= 2$  and $ \Delta_2=0.2$. Different values of 
	 of the gap outside quantum dot  $\Delta_1$ are considered. Here, (a): $\Delta_2=0.2$  
	 and (b): $\Delta_2=0.7$.}}\label{f7}
\end{figure}

 	We show the influence of  the gap outside  quantum dot $\Delta_1$  on the scattering efficiency in Figure \ref{f7}. It is clearly seen that 
 $Q$ presents  different behavior compared to	that for 
 the energy gap  $\Delta_2$ inside the quantum dot (Figure  \ref{f6}). As a result, we observe  that
 $Q$ is showing maximum values  only for small energies 
 and gaps  $\Delta_{1}$. In contrast, when  
 $ E $ increases    $ Q $ becomes mostly weak and it goes towards to a constant regardless the values taken by $\Delta_{1}$. Another remark is that the behavior of $ Q $ decreases as long as   $\Delta_{2}$ increases. 
                                              
In Figure \ref{f8} we plot the square modulus of the scattering coefficients $|c_m|^{2}$ ($ m = 0, 1, 2, 3 $)
as a function of the incident energy $ E $ {{for $V=1$, $\Delta_{2}=0.2$   and different sizes of quantum dot, with (a): $\Delta_{1}=0.1$
and (b): 	$\Delta_{1}=0.9 $.	
As a result, we observe that 
$|c_m|^{2}$ strongly depends on the size of  quantum dots. It shows  oscillatory behaviors with different amplitudes  as long as $r_0$ increase.
 For $E< \Delta_{1}$ and   small $r_0$,  $|c_m|^{2}$   decrease and increase for  large $r_0$. At the point $E=\Delta_{1} $, $|c_m|^{2}$ show minimum peaks. Compared to the results for one energy gap inside the quantum dots \cite{Jellal18}, we stress that the presence of a gap outside 
gives a non-null  square modulus of the diffusion coefficient  for both cases $E=0$ and  $E<\Delta_{1}$. Just after $E=\Delta_{1}$, it is clearly seen that 
$|c_m|^{2}$ show oscillatory behaviors.

\begin{figure}[H]
	\centering
	{\includegraphics[scale=0.65]{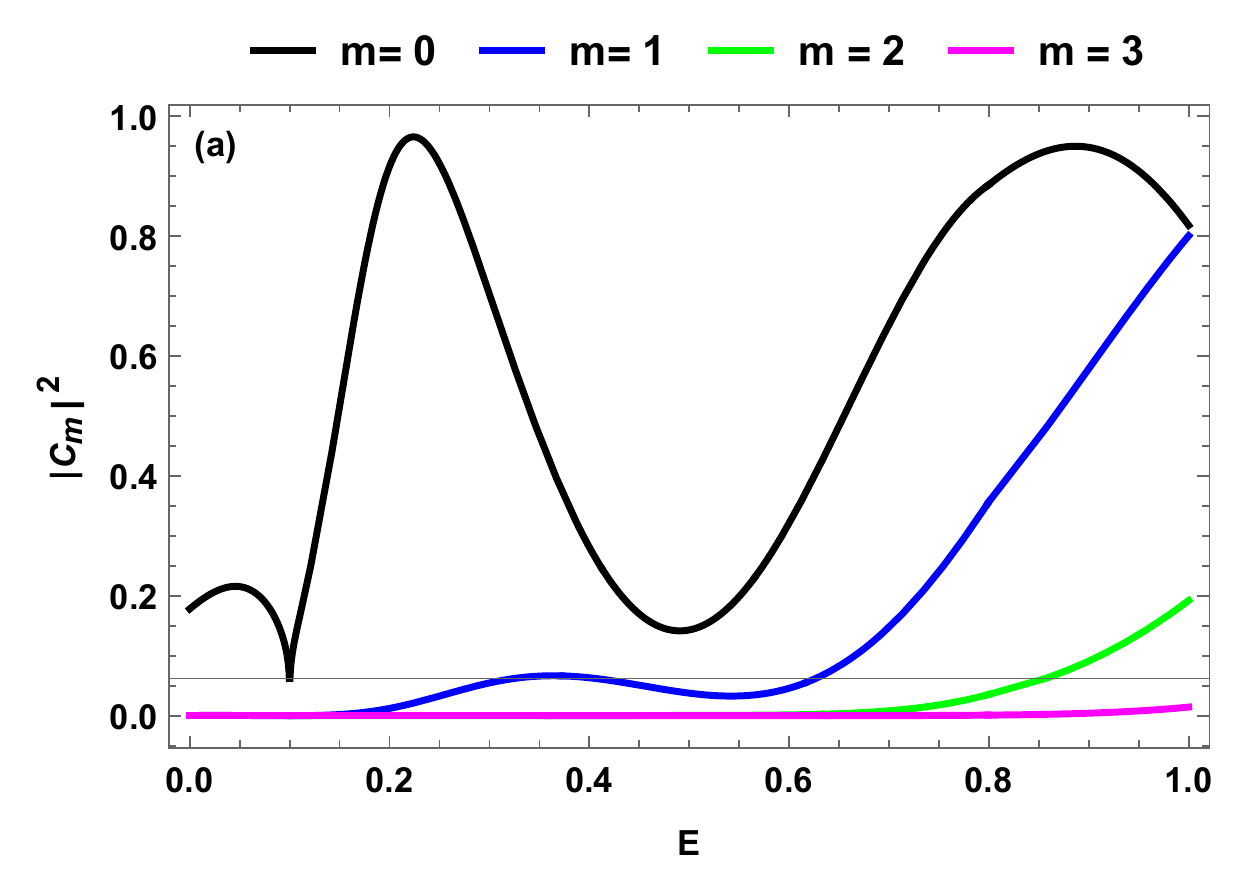}} 
	{\includegraphics[scale=0.65]{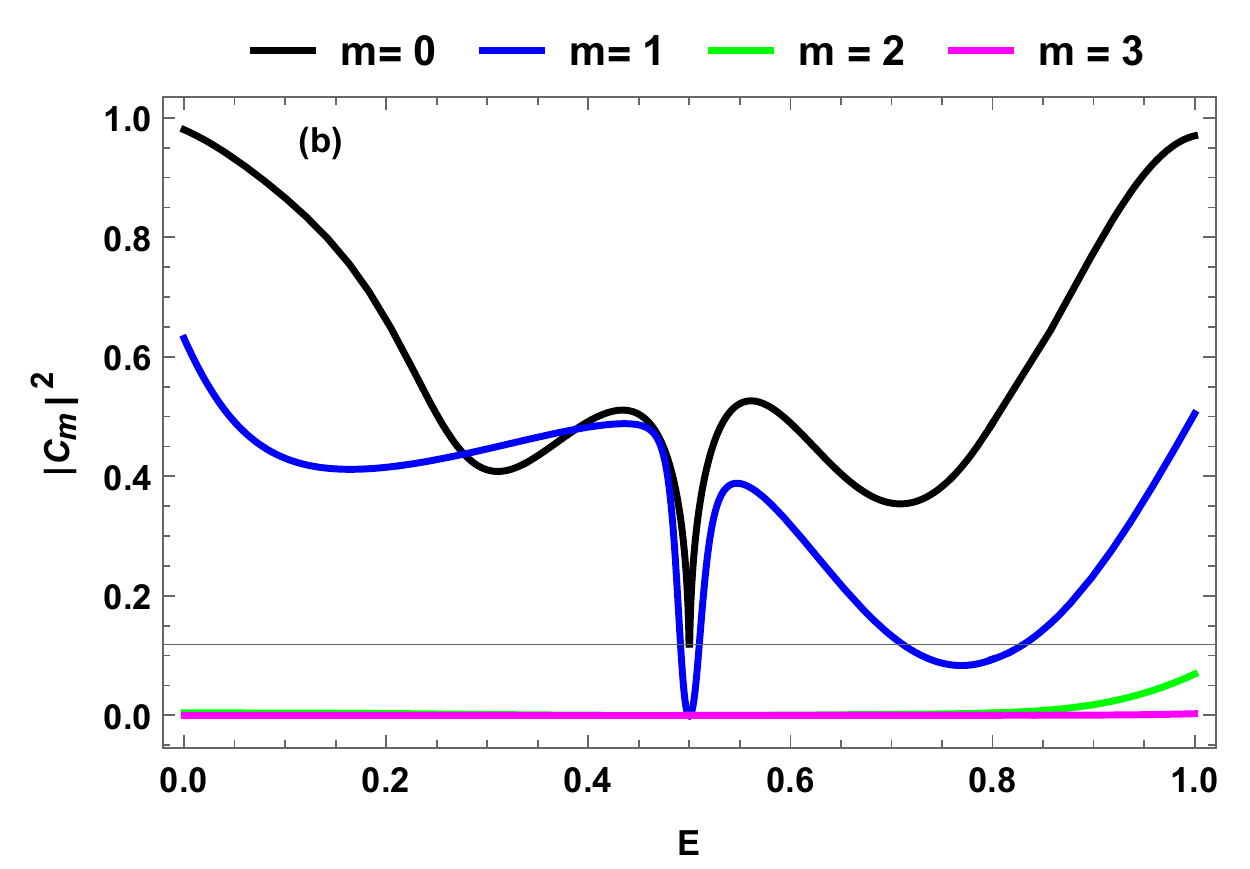}}\\
	{\includegraphics[scale=0.65]{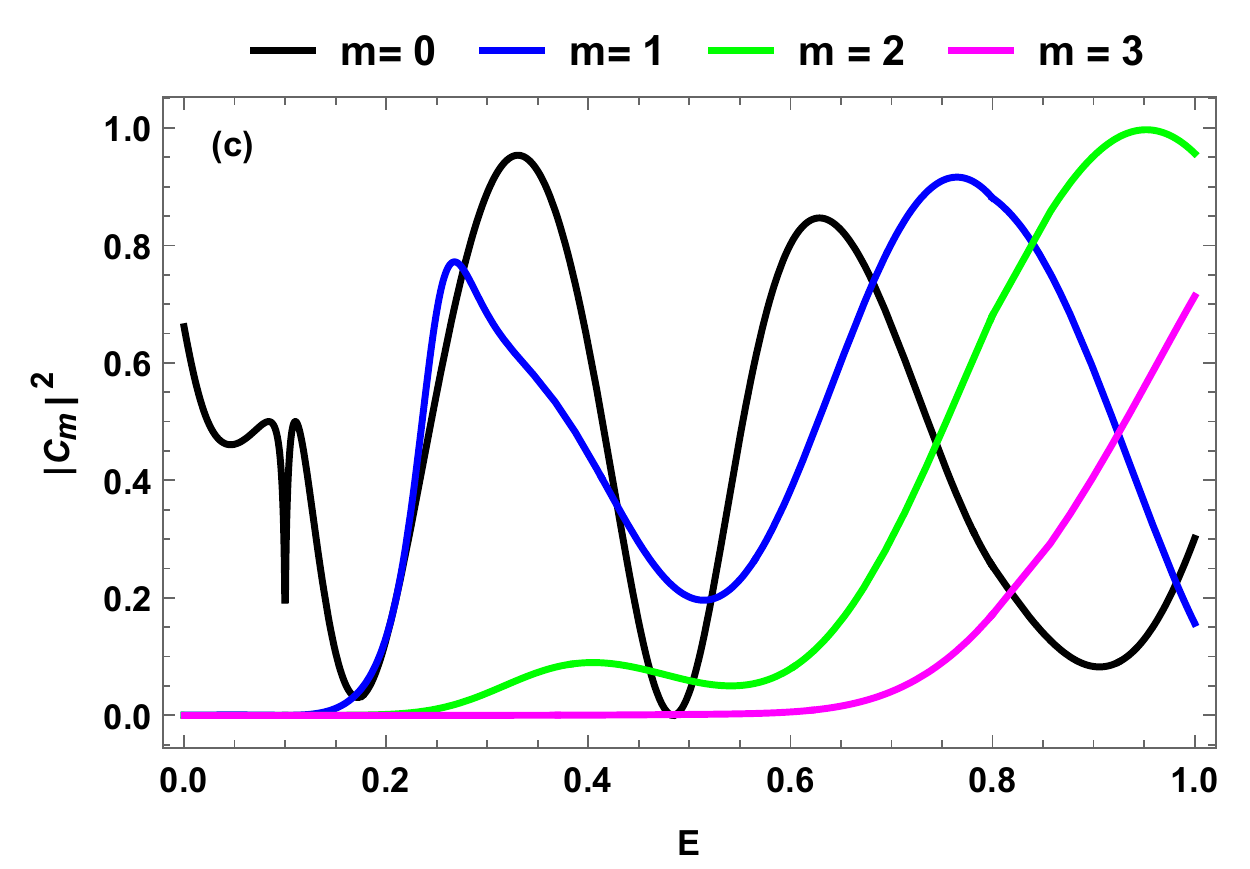}}
	{\includegraphics[scale=0.65]{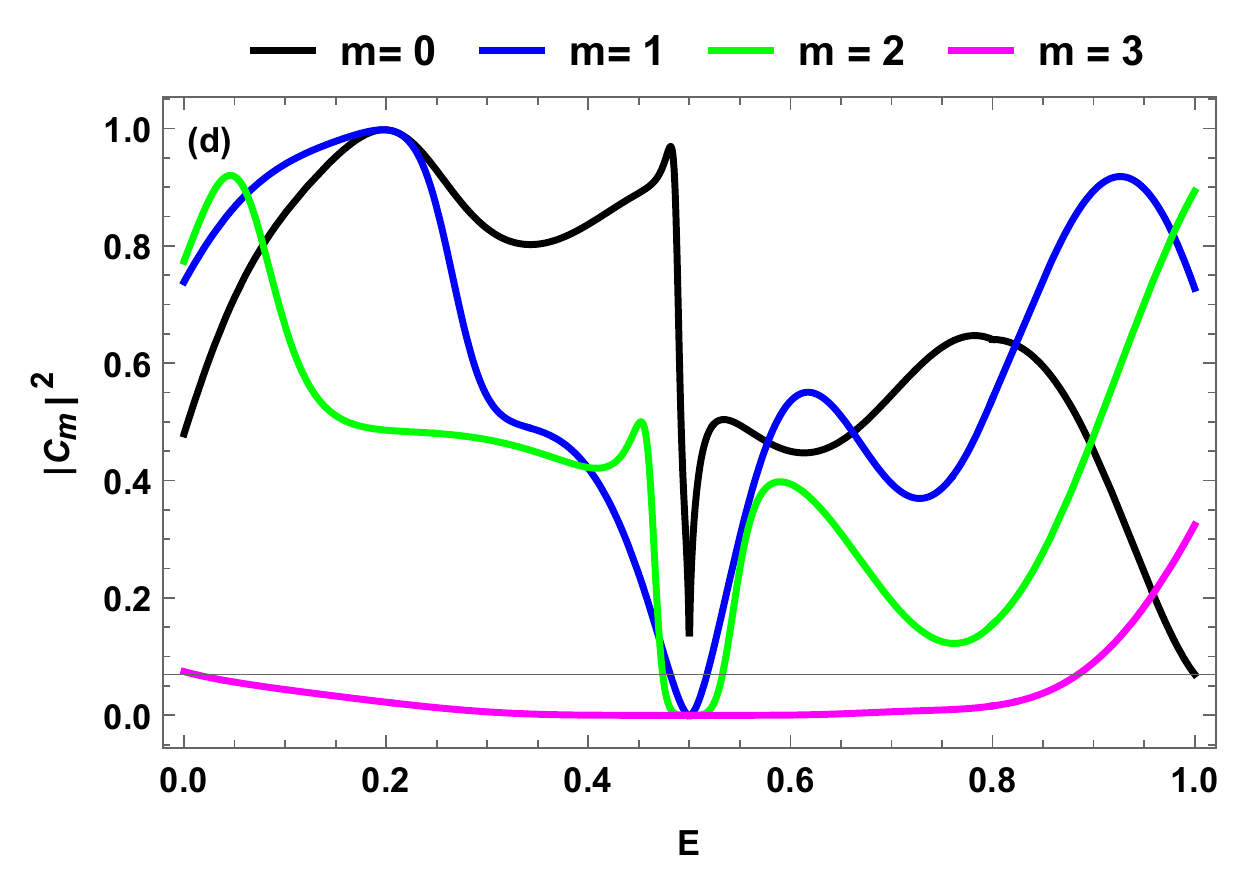}}\\
	{\includegraphics[scale=0.65]{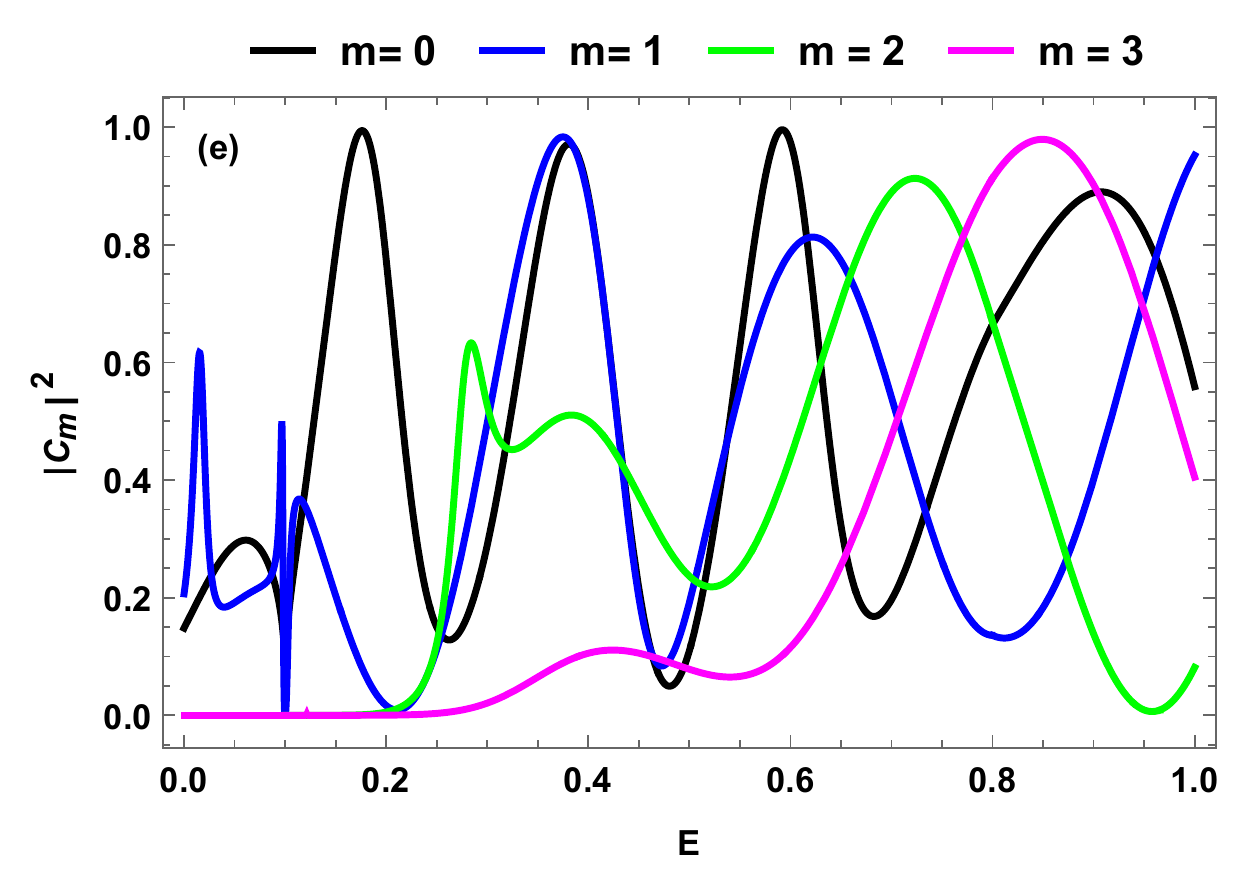}} 
	{\includegraphics[scale=0.65]{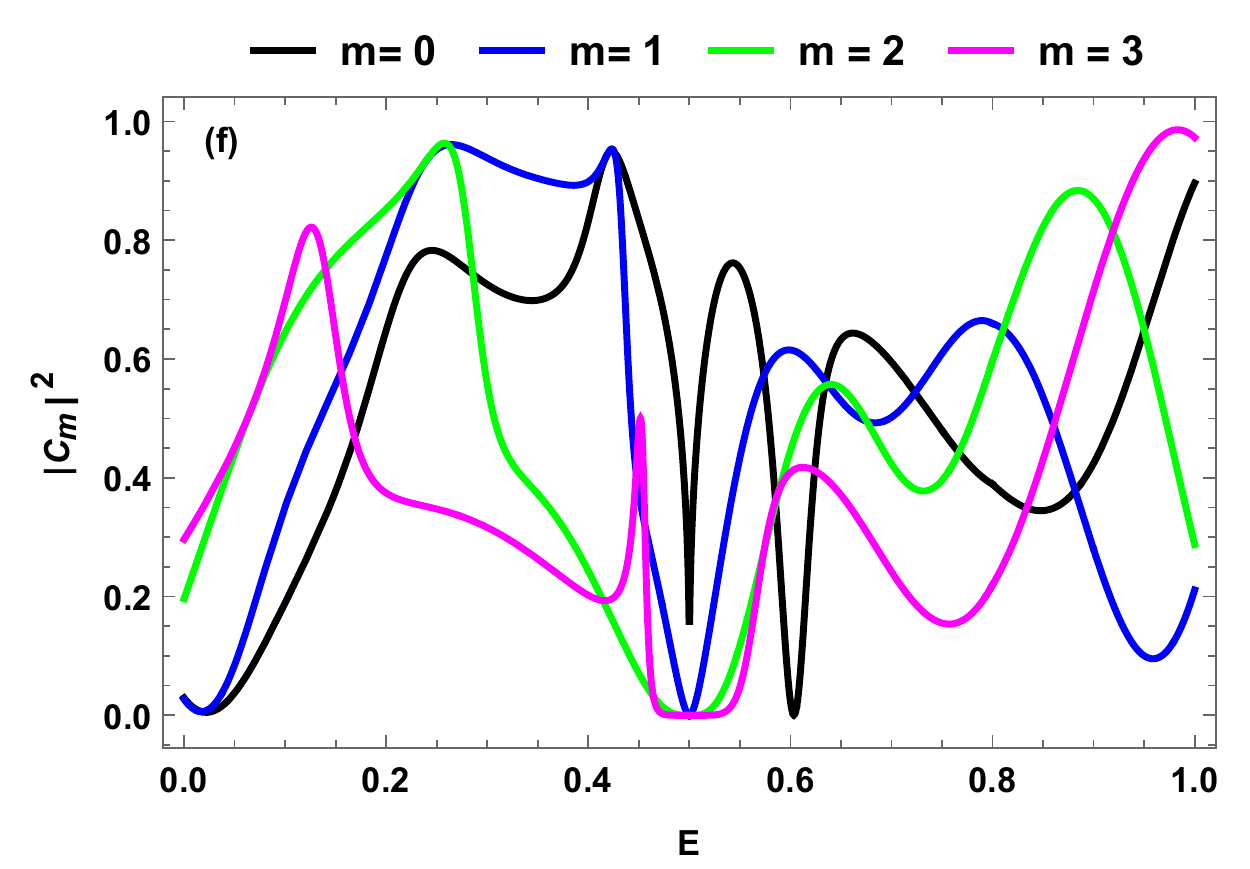}}
	\caption{{(color online) The square modulus of the scattering coefficients $|c_m|^{2}$ ($ m=0,1,2,3 $) versus  the incident energy $ E $ for $\Delta_{2}=0.2$ and $V=1$ with different sizes of quantum dot
			 (a,b): $r_0$ =3,
			 (c,d): $r_0$ =5,  
			 (e,f): $r_0$ =7. 
			  Here (a,b,c): 
			 $\Delta_1 =0.1 $ and (d,e,f): $\Delta_{1}=0.5$.}}\label{f8}
\end{figure}

 In Figure \ref{f9} we represent 
the  radial component of the reflected density current  $j_{r}^{r}$ as a function of the incident angle $\theta$ {{for $V=1$, $\Delta_{2}=0.2$ and different values of $\Delta_{1}$}}. 
We observe that  $j_{r}^{r}$ shows {{periodic oscillations with amplitudes depending on  $\Delta_{1}$.
	As a  result one sees that 
 $\Delta_1$ acts by minimizing and maximizing 
 $j_{r}^{r}$ under various choice of the physical parameters.
 For the mode $m=0$ in Figure \ref{f9}a there is a single oscillation of one maximum at $\theta=0$. For $m=1$ in Figure \ref{f9}b  there are three maximum scattering amplitudes. For $m=2$ in Figure \ref{f9}c there are five maximum scattering amplitudes and for $m=3$ in Figure \ref{f9}d there are seven maximum scattering amplitudes. We notice that the results are in agreement with those obtained for a circular  electrostatically defined quantum dot gapped in  graphene \cite{Jellal18}. In general, each mode has $ (2m+1) $ maximum scattering  observable but with the same amplitudes, that can be modified by tuning on $\Delta_1$.  

\begin{figure}[H]
	\centering
	{\includegraphics[scale=0.65]{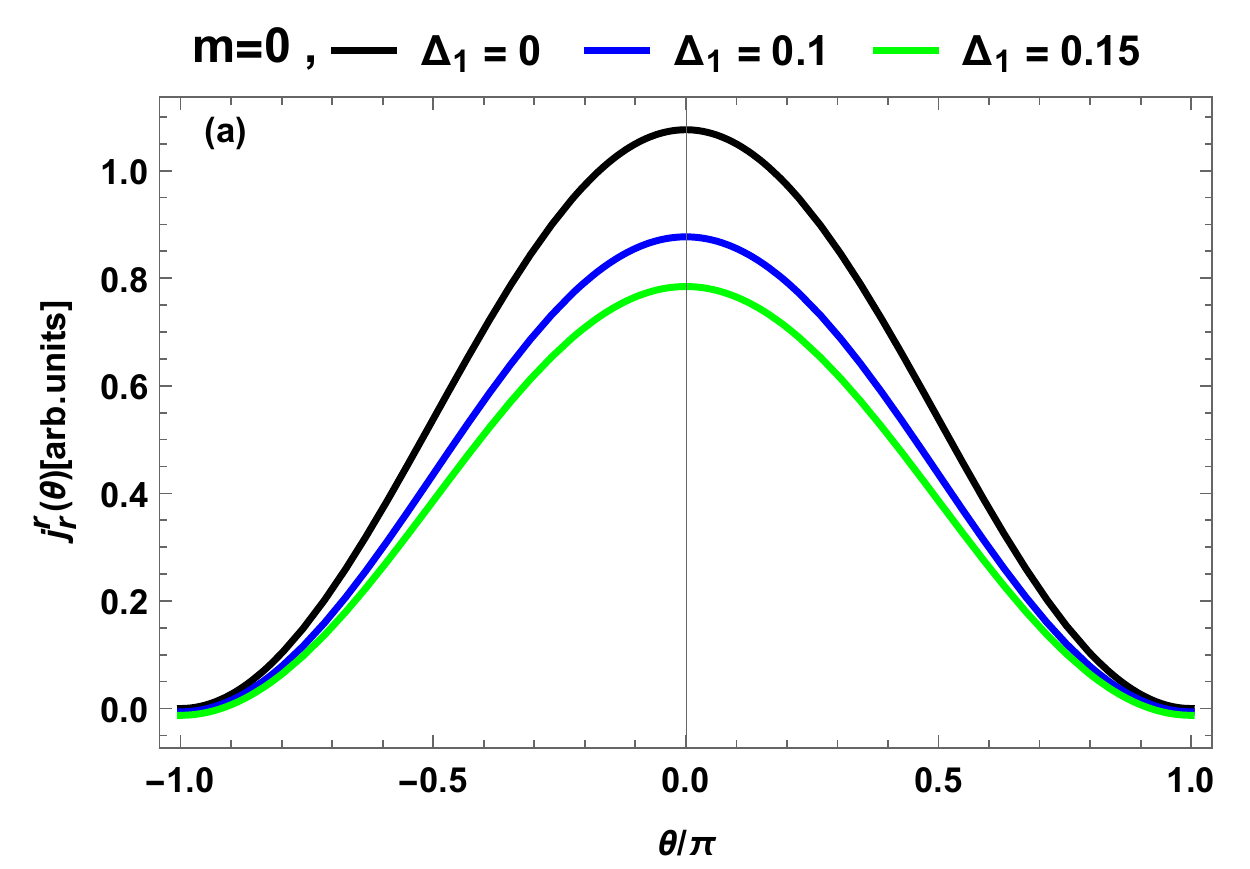}}
	{\includegraphics[scale=0.65]{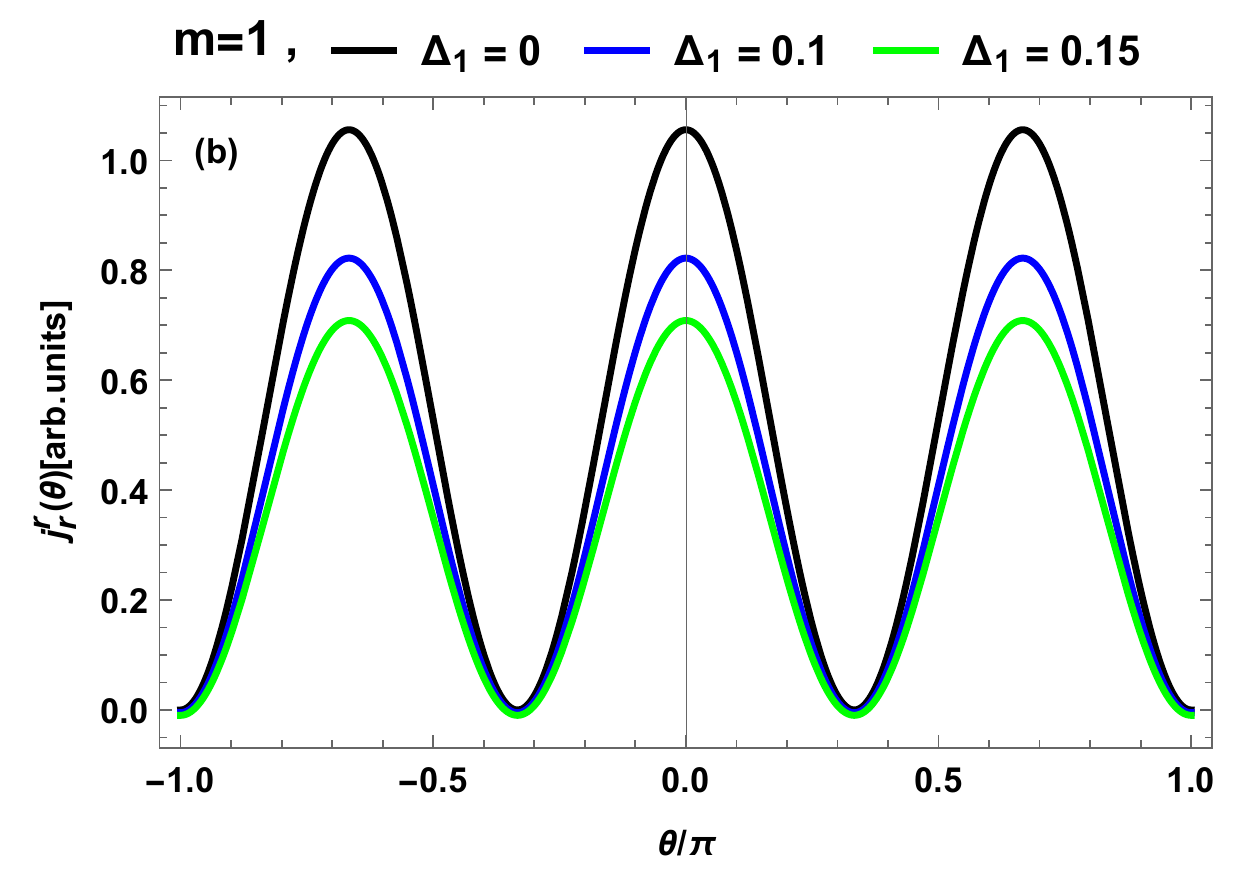}}\\
	{\includegraphics[scale=0.65]{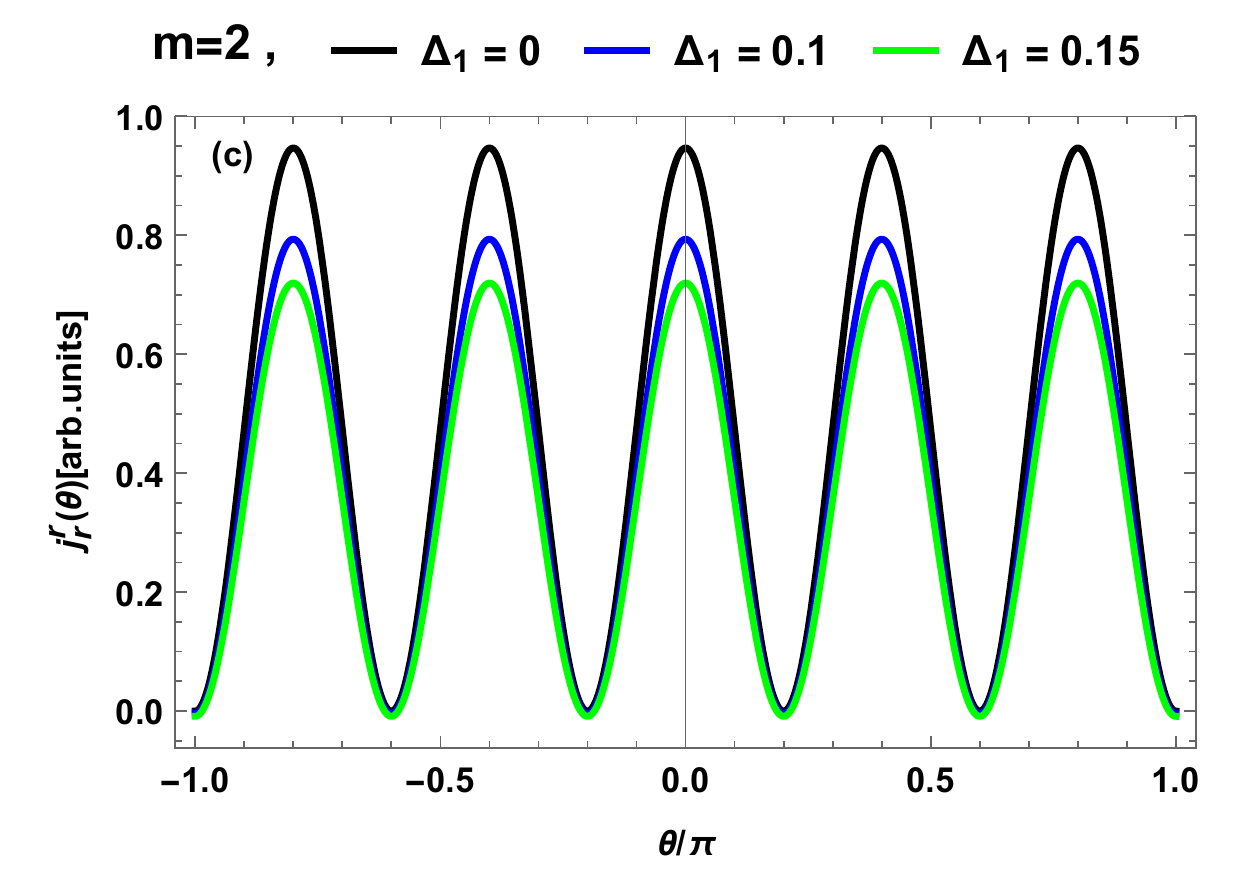}}
	{\includegraphics[scale=0.65]{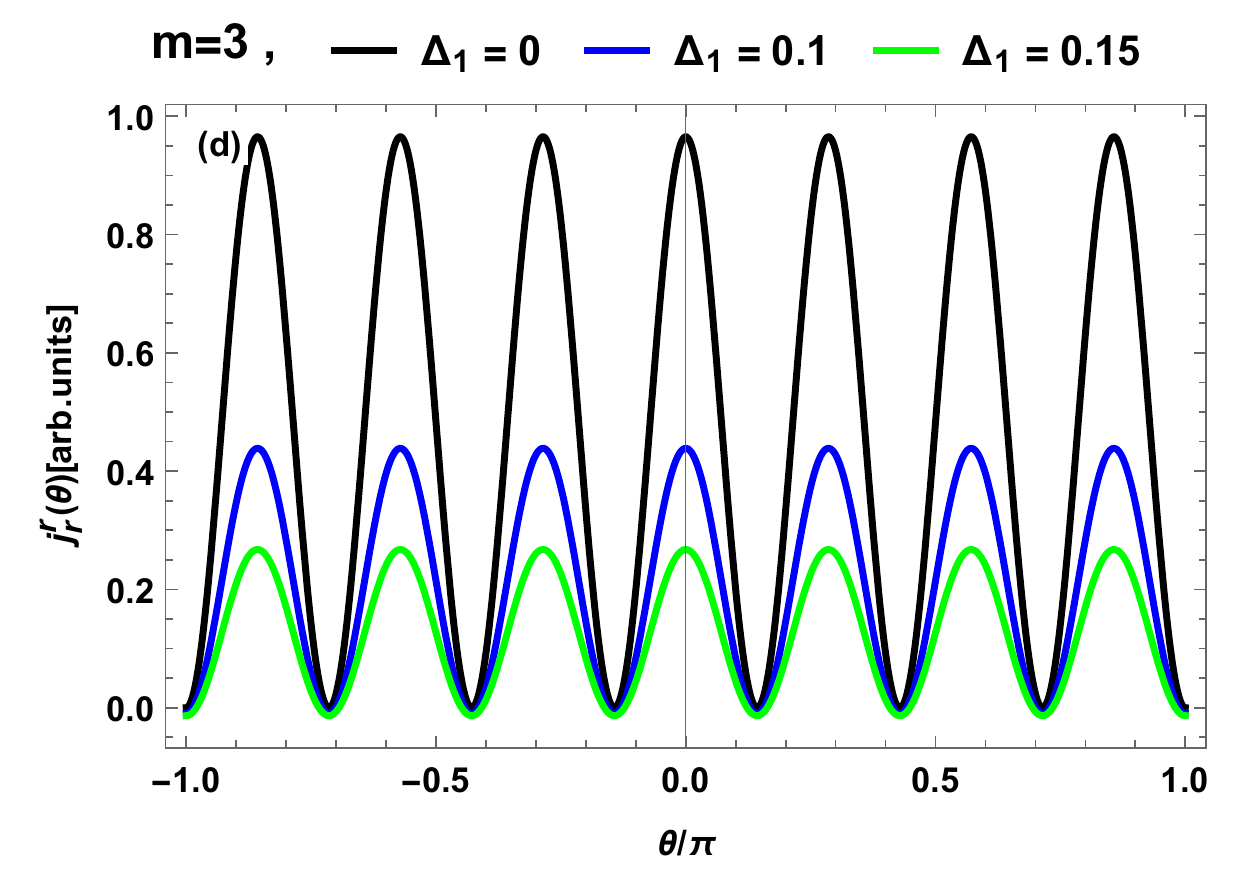}}
	\caption{{(color online) The radial component of the reflected density current $ j_r^{r} $
			versus the incident angle $ \theta $ for $\Delta_2 =0.2 $   and $V=1 $, (a): $ E=0.6 $  and  $r_0=3.3$, (b): $ E=0.65$  and $ r_0= 5$, (c): $E=0.69$ and $ r_0 =7.25 $, (d): $ E =0.35$ and $ r_0=8.7$. Different values of the gap outside quantum dot $\Delta_1$ are considered.}}\label{f9}
\end{figure}

\section{Conclusion}

We have studied the scattering of Dirac electrons in circular graphene quantum dot through an electrostatic  potential
 $ V $  with  mass-inverted  terms  $\Delta_{1}$ outside and
 $\Delta_2$
 inside. 
The solutions of energy spectrum are  found to be dependent on both gaps 
$\Delta_{1}$ and $\Delta_2$.
Using the boundary condition at the interface, the scattering coefficients
are explicitly determined. By focusing on solutions at large arguments we have used the asymptotic behavior of Hankel function to obtain analytical expressions of
the  radial component $j^r_r$ of  
current density associated to the reflected wave, 
 the scattering efficiency $Q$ and the square modulus of the scattering coefficient $|c_m|^{2}$.

Our numerical analysis showed that $Q$ can be controlled by tuning
the energy gap $\Delta_{1}$ outside the quantum  dots. More precisely, 
 we have observed that $Q$ decreases as long as $\Delta_{1}$ increases. Furthermore, $Q$  shows  oscillatory behaviors for some chosen values of $\Delta_{1}$. Also
among the obtained new results, we have observed  that 
$\Delta_{1}$ strongly affects the behavior of $|c_m|^{2}$ as a function 
of the incident energy $E$. Indeed, 
$|c_m|^{2}$ is non-null for two  energies  $E=0$ and  $E<\Delta_{1}$ contrary to 
the case of one energy gap inside the quantum dots \cite{Jellal18}.
For the radial component of the reflected  current it was showed that each mode has $ (2m+1) $ maximum scattering directions observable but with same amplitudes, which can be controlled under the turn on  of $\Delta_{1}$.

\end{document}